\documentclass[11pt]{article}
\usepackage{a4p}
\usepackage{amssymb}
\usepackage{epsfig}
\usepackage{epsf}
\usepackage{cite}
\usepackage{pennames}
\usepackage{rotating}
\usepackage{array}
\newcommand{\inmath}[1] {\ifmmode#1\else$#1$\fi}
\newcommand{\definmath}[2] {\def#1{\ifmmode#2\else$#2$\fi}}

\definmath{\PWpm} {\mathrm{W}^{\pm}}      % W+-
\definmath{\Plp} {\ell^{+}}        % l+
\definmath{\Plm} {\ell^{-}}        % l-
\definmath{\Plpm}   {\ell^{\pm}}         % l+-
\definmath{\Pgtp} {\tau^{+}}        % tau+
\definmath{\Pgtm} {\tau^{-}}        % tau-
\definmath{\Pgtpm}   {\tau^{\pm}}         % tau+-
\definmath{\Pgn}  {\nu}          % neutrino
\definmath{\Pagn} {\overline{\nu}}     % anti-neutrino
\definmath{\Pf}      {\mathrm{f}}
\definmath{\Paf}  {\overline{\mathrm{f}}}
\definmath{\Pq}      {\mathrm{q}}
\definmath{\Paq}  {\overline{\mathrm{q}}}
\definmath{\Pu}      {\mathrm{u}}
\definmath{\Pau}  {\overline{\mathrm{u}}}
\definmath{\Pd}      {\mathrm{d}}
\definmath{\Pad}  {\overline{\mathrm{d}}}
\definmath{\Ps}      {\mathrm{s}}
\definmath{\Pas}  {\overline{\mathrm{s}}}
\definmath{\Pc}      {\mathrm{c}}
\definmath{\Pac}  {\overline{\mathrm{c}}}
\definmath{\Pb}      {\mathrm{b}}
\definmath{\Pab}  {\overline{\mathrm{b}}}
\definmath{\Pt}      {\mathrm{t}}
\definmath{\Pat}  {\overline{\mathrm{t}}}
\definmath{\Pap}  {\overline{\mathrm{p}}}
\definmath{\Pan}  {\overline{\mathrm{n}}}
\definmath{\PaD}  {\overline{\mathrm{D}}}
\definmath{\PaDz} {\overline{\mathrm{D}}^{0}}
\definmath{\PaB}  {\overline{\mathrm{B}}}
\definmath{\PaBz} {\overline{\mathrm{B}}^{0}}
\definmath{\PsDpm}   {\mathrm{D}^{\pm}_{\mathrm{s}}}  % Ds+-
\definmath{\PcgLpm}  {\Lambda^{\pm}_{\mathrm{c}}}  % Lambda_c+-
\definmath{\PD} {\mathrm{D}}     % D
\definmath{\PDst} {\mathrm{D}^{*}}     % D*
\definmath{\PgLz} {\Lambda^{0}}        % Lambda0

\newcommand{\epem}   {\Pep\Pem}

\newcommand{\nunug}   {\Pgn\Pagn$\mathrm{\gamma}~$}

\newcommand{\nunu}   {\Pgn\Pagn}
\newcommand{\qqbar}  {\Pq\Paq}
\newcommand{\qqbarg}  {\Pq\Paq$\mathrm{\gamma}~$}

\def\lapproxeq {\mbox{{\lower .7ex\hbox{$\;\stackrel{\textstyle
                  <}{\sim}\;$}}}}
\def\gapproxeq  {\mbox{{\lower .7ex\hbox{$\;\stackrel{\textstyle
                  >}{\sim}\;$}}}}

\definmath{\GeV}  {\mathrm{GeV}}
\definmath{\GeVc} {\mathrm{GeV}\!/c}
\definmath{\GeVcc}   {\mathrm{GeV}\!/c^2}
\definmath{\MeV}  {\mathrm{MeV}}
\definmath{\MeVc} {\mathrm{MeV}\!/c}
\definmath{\MeVcc}   {\mathrm{MeV}\!/c^2}
\definmath{\MVm}  {\mathrm{MV}\!/\mathrm{m}}
\definmath{\keV}  {\mathrm{keV}}
\definmath{\keVcm}   {\mathrm{keV}\!/\mathrm{cm}}
\definmath{\kV}      {\mathrm{kV}}
\definmath{\km}      {\mathrm{km}}
\definmath{\meter}   {\mathrm{m}}
\definmath{\cm}      {\mathrm{cm}}
\definmath{\mm}      {\mathrm{mm}}
\definmath{\micron}  {\mu\mathrm{m}}
\definmath{\nm}      {\mathrm{nm}}
\definmath{\kg}      {\mathrm{kg}}
\definmath{\gram} {\mathrm{g}}
\definmath{\second}  {\mathrm{s}}
\definmath{\microsec}   {\mu\mathrm{s}}
\definmath{\degree}  {^\circ}
\definmath{\degC} {^\circ\mathrm{C}}
\definmath{\ohm}  {\Omega}
\definmath{\Mohm} {\mathrm{M}\Omega}
\definmath{\rad}  {\mathrm{rad}}
\definmath{\mrad} {\mathrm{mrad}}
\definmath{\nb}      {\mathrm{nb}}
\newcommand{\eqref}[1]  {(\ref{#1})}

\newcommand{\PRL}[3]  {Phys.~Rev.\ Lett.\ \textbf{#1} (#2) #3}

\newcommand{\PhysRev}   {Phys.~Rev.}
\newcommand{\NPhys}  {Nucl.~Phys.}
\newcommand{\NIM} {Nucl.~Instr.\ Meth.}
\newcommand{\ZPhys}  {Z.~Phys.}
\newcommand{\IEEENS} {IEEE Trans.\ Nucl.~Sci.}
\newcommand{\CPC} {Comp. Phys. Comm.}
\newcommand{\EPJ} {Eur.~Phys.~J.}

\newcommand{\OPALColl}    {OPAL Collab.}
\newcolumntype{L} {>{$}l<{$}}
\newcolumntype{C} {>{$}c<{$}}
\newcolumntype{R} {>{$}r<{$}}

\def\etal{\mbox{{\it et al.}}}
\newcommand{\EPC}[3]  {Eur.\ Phys.\ J.\ \textbf{C#1} (#2) #3}
\newcommand{\PLB}[3]  {Phys.\ Lett.\ \textbf{B#1} (#2) #3}
\newcommand{\ZPC}[3]  {Z.\ Phys.\ \textbf{C#1} (#2) #3}

\newcommand{\PRD}[3]  {Phys.\ Rev.\ \textbf{D#1} (#2) #3}
\newcommand{\NPB}[3]  {Nucl.\ Phys.\ \textbf{B#1} (#2) #3}
\def\opalabbien{OPAL Collaboration, G.\ Abbiendi \etal}

\newcommand{\Jetset}{\mbox{J{\sc etset}}}

\newcommand{\KORALW}{\mbox{K{\sc oralw}}}
\newcommand{\KORALZ}{\mbox{K{\sc oralz}}}
\newcommand{\Excalibur}{\mbox{E{\sc xcalibur}}}

\newcommand{\grcff}{\mbox{grc4f}}
\newcommand{\Pythia}{\mbox{P{\sc ythia}}}

\newcommand{\kk}{\mbox{KK2f}}

\newcommand{\NUNUGPV}{\mbox{N{\sc unugpv98}}}

\newcommand{\BHWIDE}{\mbox{B{\sc hwide}}}
\newcommand{\PHOJET}{\mbox{P{\sc hojet}}}

\newcommand{\Herwig}{\mbox{H{\sc erwig}}}
\newcommand{\HERWIG}{\mbox{H{\sc erwig}}}

\definmath{\Pq}      {\mathrm{q}}
\definmath{\Paq}  {\overline{\mathrm{q}}}
\definmath{\Pu}      {\mathrm{u}}
\definmath{\Pau}  {\overline{\mathrm{u}}}
\definmath{\Pd}      {\mathrm{d}}
\definmath{\Pad}  {\overline{\mathrm{d}}}
\definmath{\Ps}      {\mathrm{s}}
\definmath{\Pas}  {\overline{\mathrm{s}}}
\definmath{\Pc}      {\mathrm{c}}
\definmath{\Pac}  {\overline{\mathrm{c}}}
\definmath{\Pb}      {\mathrm{b}}
\definmath{\Pab}  {\overline{\mathrm{b}}}
\definmath{\Pt}      {\mathrm{t}}
\definmath{\Pat}  {\overline{\mathrm{t}}}
\definmath{\Pgtp} {\tau^{+}}        % tau+
\definmath{\Pgtm} {\tau^{-}}        % tau-
\definmath{\Pgtpm}   {\tau^{\pm}}         % tau+-

\newcommand{\stat}   {\ensuremath{\mathrm{(stat.)}}}
\newcommand{\syst}   {\ensuremath{\mathrm{(syst.)}}}

\def\etal{\mbox{{\it et al.}}}

\def\gappeq{\ensuremath{\mathrel{ \rlap{\raise.5ex\hbox{>}}
                      {\lower.5ex\hbox{\sim}}}}}
\def\lappeq{\ensuremath{\mathrel{ \rlap{\raise.5ex\hbox{<}}
                      {\lower.5ex\hbox{\sim}}}}}
\definmath{\GeV}  {\mathrm{GeV}}
\definmath{\GeVc} {\mathrm{GeV}\!/c}
\definmath{\GeVcc}   {\mathrm{GeV}\!/c^2}
\definmath{\MeV}  {\mathrm{MeV}}
\definmath{\MeVc} {\mathrm{MeV}\!/c}
\definmath{\MeVcc}   {\mathrm{MeV}\!/c^2}
\definmath{\MVm}  {\mathrm{MV}\!/\mathrm{m}}
\definmath{\keV}  {\mathrm{keV}}
\definmath{\keVcm}   {\mathrm{keV}\!/\mathrm{cm}}
\definmath{\kV}      {\mathrm{kV}}
\definmath{\km}      {\mathrm{km}}
\definmath{\meter}   {\mathrm{m}}
\definmath{\cm}      {\mathrm{cm}}
\definmath{\mm}      {\mathrm{mm}}
\definmath{\micron}  {\mu\mathrm{m}}
\definmath{\nm}      {\mathrm{nm}}
\definmath{\kg}      {\mathrm{kg}}
\definmath{\gram} {\mathrm{g}}
\definmath{\second}  {\mathrm{s}}
\definmath{\microsec}   {\mu\mathrm{s}}
\definmath{\degree}  {^\circ}
\definmath{\degC} {^\circ\mathrm{C}}
\definmath{\ohm}  {\Omega}
\definmath{\Mohm} {\mathrm{M}\Omega}
\definmath{\rad}  {\mathrm{rad}}
\definmath{\mrad} {\mathrm{mrad}}
\definmath{\nb}      {\mathrm{nb}}
\definmath{\pb}      {\mathrm{pb}}
\begin{document}
\begin{titlepage}
\begin{center}{\large   EUROPEAN ORGANIZATION FOR NUCLEAR RESEARCH
}\end{center}\bigskip
\begin{flushright}
  CERN-EP-2000-067   \\ 
   30 May 2000 \\
%  OPAL PR311 \\ 
%  26 May 2000
\end{flushright}
\bigskip\bigskip\bigskip
\begin{center}
 {\huge\bf \boldmath
   Search for Trilinear Neutral \\ Gauge Boson Couplings \\
   in Z$\gamma$ production at $\sqrt{s}=189$~GeV at LEP}
\end{center}
\bigskip\bigskip
\bigskip\bigskip
\begin{center}{\LARGE The OPAL Collaboration
}\end{center}
\vspace{0.5cm}
%\begin{center}   {Authors: S.~Arcelli, S.~Spagnolo}
%\end{center}
%\begin{center}   
%{Editorial Board: H.~Rick, 
%D.~Strom, C.P.~Ward and  G.W.~Wilson}
%\end{center}
\bigskip\bigskip \bigskip
\begin{center}{\large Abstract}\end{center}
{The data recorded at a centre-of-mass energy of 189~\GeV\ by the OPAL
detector at LEP are used to search for trilinear 
couplings of the neutral gauge bosons in the process 
\epem$\rightarrow{\rm Z\gamma}$. 
The cross-sections for multihadronic events with an 
energetic isolated photon, and for events with a high energy photon
accompanied  by missing energy are measured. These cross-sections and 
the photon energy, polar angle and isolation angle distributions 
are compared to the Standard Model predictions 
and to the theoretical expectations of a model 
allowing for ${\rm Z\gamma Z}$ and ${\rm Z\gamma\gamma}$ 
vertices. 
Since no significant deviations with respect 
to the Standard Model expectations are found, 
constraints are derived on the strength of neutral trilinear gauge 
couplings.}
\bigskip\bigskip
\bigskip\bigskip
%\begin{center}   {Please send comments to 
%Stefania.Spagnolo@cern.ch or Silvia.Arcelli@cern.ch 
%by 25 April 2000.}
%\end{center}
\vspace{0.5cm}
%\bigskip\bigskip\bigskip\bigskip
\bigskip\bigskip
\begin{center}{\large
  (Submitted to Eur. Phys. J.)
}\end{center}
\bigskip\bigskip
\bigskip\bigskip
\end{titlepage}
%%%%%%%%%%%%%%%%%%%%%%%%%%%%%%%%%%%%%%%%%%%%%%%%%%%%%%%%%%%%%%%%%%%%%
%%%%%%%%%%%%%%%%%%%%%%%%%%%%%%%%%%%%%%%%%%%%%%%%%%%%%%%%%%%%%%%%%%%%%
%%%%%%%%%%%%%%%%%%%%%%%%%%%%%%%%%%%%%%%%%%%%%%%%%%%%%%%%%%%%%%%%%%%%%
%\input{authorlist.tex}
\begin{center}{\Large        The OPAL Collaboration
}\end{center}\bigskip
\begin{center}{
%begin authorlist PLEASE DO NOT DELETE THIS COMMENT
G.\thinspace Abbiendi$^{  2}$,
K.\thinspace Ackerstaff$^{  8}$,
C.\thinspace Ainsley$^{  5}$,
P.F.\thinspace Akesson$^{  3}$,
G.\thinspace Alexander$^{ 22}$,
J.\thinspace Allison$^{ 16}$,
K.J.\thinspace Anderson$^{  9}$,
S.\thinspace Arcelli$^{ 17}$,
S.\thinspace Asai$^{ 23}$,
S.F.\thinspace Ashby$^{  1}$
% ...
%end authorlist PLEASE DO NOT DELETE THIS COMMENT
}\end{center}\bigskip
\bigskip
%begin institutes
$^{  1}$School of Physics and Astronomy, University of Birmingham,
Birmingham B15 2TT, UK
\newline
$^{  2}$Dipartimento di Fisica dell' Universit\`a di Bologna and INFN,
I-40126 Bologna, Italy
\newline
$^{  3}$Physikalisches Institut, Universit\"at Bonn,
D-53115 Bonn, Germany
\newline
% ...
$^{  5}$Cavendish Laboratory, Cambridge CB3 0HE, UK
\newline
% ...
$^{  8}$CERN, European Organisation for Nuclear Research,
CH-1211 Geneva 23, Switzerland
\newline
$^{  9}$Enrico Fermi Institute and Department of Physics,
University of Chicago, Chicago IL 60637, USA
\newline
% ...
$^{ 16}$Department of Physics, Schuster Laboratory, The University,
Manchester M13 9PL, UK
\newline
$^{ 17}$Department of Physics, University of Maryland,
College Park, MD 20742, USA
\newline
% ...
$^{ 22}$Department of Physics and Astronomy, Tel Aviv University,
Tel Aviv 69978, Israel
\newline
$^{ 23}$International Centre for Elementary Particle Physics and
Department of Physics, University of Tokyo, Tokyo 113-0033, and
Kobe University, Kobe 657-8501, Japan
\newline
% ...
Budapest, Hungary.
%end notes
%%%%%%%%%%%%%%%%%%%%%%%%%%%%%%%%%%%%%%%%%%%%%%%%%%%%%%%%%%%%%%%%%%%%%
%%%%%%%%%%%%%%%%%%%%%%%%%%%%%%%%%%%%%%%%%%%%%%%%%%%%%%%%%%%%%%%%%%%%%
%%%%%%%%%%%%%%%%%%%%%%%%%%%%%%%%%%%%%%%%%%%%%%%%%%%%%%%%%%%%%%%%%%%%%

\clearpage\newpage
%-----------------------------------------------------------------------
\section{Introduction}             \label{sec:intro}
%-----------------------------------------------------------------------

The self-interactions of the gauge bosons are consequences 
of the non-Abelian structure of the electroweak sector of the Standard 
Model; therefore, the strength of trilinear and quartic gauge couplings is 
predicted as a result of the gauge symmetry of the theory.
The study of trilinear gauge boson couplings in
two-boson production processes is within the reach of existing
accelerators and measurements of WWZ and WW$\gamma$ couplings are being 
performed with increasing precision 
in \epem\ \cite{bib:lepagc} and $\rm{p \bar p}$ \cite{bib:tevagc} collisions. 
While non-zero values are predicted for the couplings in the triple  
and quartic vertices involving charged gauge bosons, the tree level
vertices Z$\gamma$Z, Z$\gamma\gamma$ and ZZZ are not generated by the
Standard Model Lagrangian;  higher order corrections through virtual
loops contribute at the level of 10$^{-4}$ \cite{bib:ren}, well below the
 current experimental sensitivity. Nevertheless, new phenomena 
with a characteristic
mass scale above the present experimental threshold
 might lead to tree-level neutral trilinear gauge couplings (NTGC) in 
the effective Lagrangian  
\cite{bib:georgi,bib:gounren93} parametrising the residual low energy 
effects from new physics.
For example, as suggested in~\cite{bib:gounren}, 
virtual effects from new heavy fermions having non-standard couplings to 
the gauge bosons might generate sizeable anomalous contributions.

The most general Z$\gamma$V vertex (where V is the
intermediate virtual boson, either photon or a Z) compatible 
with Lorentz invariance and electro-magnetic gauge invariance involves four 
independent operators, corresponding to the allowed helicity states 
for the Z$\gamma$ pair \cite{bib:gaemers,bib:hagiwara}. Therefore, in a model
independent description, there exist eight couplings: four of them (${\rm
  h_{\it i}^Z}$, $i=1,...,4$) corresponding to V=Z and four (${\rm
  h_{\it i}^\gamma}$ ) corresponding to V=$\gamma$.
The vertex function was first given in \cite{bib:hagiwara}. In this
analysis, we adopt the most recent convention established in 
\cite{bib:gounren}. 
The lowest dimensional operators associated to the 
${\rm h_{\it i}^{Z,\gamma}}$ ($i=1,3$) couplings are of 
dimension six, while dimension eight operators are associated to 
${\rm h_{\it i}^{Z,\gamma}}$ ($i=2,4$). As discussed in~\cite{bib:gounren}, 
 no additional symmetry constraints, such as the SU(2)$\times$U(1) 
 gauge invariance 
%of the Effective Lagrangian 
usually assumed in the case of WWV anomalous couplings 
\cite{bib:su2u1}, can help in reducing the number of free parameters in the 
neutral gauge vertex, since operators of even higher dimensionality 
would be required. 
Therefore a model independent approach is used which retains all eight 
couplings.

In this paper, the process \epem$\rightarrow \rm Z\gamma$ at 
$\sqrt{s}=189$~GeV is investigated through the final states 
\qqbarg and $\mathrm{\nu \bar \nu \gamma}$, the dominant 
decay modes of the $\rm{Z}$ boson,  
with the aim of searching for Z$\gamma$Z and 
Z$\gamma\gamma$ couplings. 
Experimental constraints on these 
couplings have been produced in the past from the analysis of LEP data 
at lower energies  \cite{bib:nagclep} and of the {\sc Tevatron} data 
\cite{bib:nagctev}. The analysis presented here has 
higher sensitivity due to the increased centre-of-mass energy 
and due to the large data sample collected during the 1998 operation 
of LEP. It also benefits from the recent clarification~\cite{bib:gounren}
of the theoretical framework in which neutral gauge 
boson self-interactions can be 
described. A recent analysis based on data collected 
at $\rm{\sqrt{s}=189~GeV}$ by the L3 
collaboration~\cite{bib:ntgcL3}  adopts the same convention
as in this paper. 
On the other hand, due to this different convention, the comparison of 
the results presented here and in~\cite{bib:ntgcL3} 
with previous published results is not straightforward.

In \epem\ collisions, 
the production of Z$\gamma$ final states via anomalous neutral gauge
couplings has a large irreducible Standard Model background from 
$\rm{Z^{0}}$ production with hard initial state radiation (ISR).  
Small contributions to \qqbarg production also arise
from \epem$\rightarrow\gamma\gamma^*\rightarrow \gamma {\rm q\bar{q}}$ 
and from final state radiation in \qqbar\ production. 
In the \nunug channel, final states arising from the exchange 
of a W boson in the t-channel also contribute to the cross-section,
although their contribution is small 
within the Z$\gamma$ signal acceptance used in this analysis.   
As a general property, the Z and $\gamma$ produced at the anomalous
vertices are more isotropically distributed than the dominant 
Standard Model background, which is characterized by
the strongly forward peaked 
angular distribution of initial state radiation. 
Therefore, deviations from the Standard Model predictions 
due to Z$\gamma$V couplings would be more pronounced for 
large angles between the beam direction and the photon. 
In the \qqbarg final state, the angular distribution of the jets can
also be exploited in order to gain sensitivity to anomalous couplings,
due to the resulting enhancement of the 
longitudinal polarisation of the Z boson 
affecting the fermion decay angle. 
On the other hand, the photon energy spectrum has a marginal
sensitivity, due to the kinematic constraints from the 
fixed centre-of-mass energy and the narrow Z resonance. 
Since all the terms in the Z$\gamma$V
vertex are proportional to the momenta of the gauge bosons involved,
this results in an enhancement of the sensitivity to NTGC 
as the centre-of-mass energy increases. 
Finally, the experimental signature 
of the anomalous ${\rm h_{\it i}^{Z,\gamma}}$
couplings depends on the CP parity of the associated operators. 
The vertex terms proportional to ${\rm h_1^{Z,\gamma}}$  
and ${\rm h_2^{Z,\gamma}}$ violate CP and, hence, do not interfere 
with the CP conserving Standard Model amplitudes. As a result the total 
and differential cross-sections receive only additive contributions 
from the anomalous processes. The remaining couplings, associated to 
CP even terms, lead to amplitudes which interfere with the
Standard Model; therefore the differential and total cross-sections are enhanced or suppressed depending on the sign and the 
size of the  ${\rm h_{\it i}^{Z,\gamma}}$ ($i=3,4$) couplings. 

%While
%this feature is responsible for useful signatures of NTGC in hadronic
%collisions, at LEP the resulting sensitivity of the photon energy spectrum
%is marginal due to the kinematic constraints from the fixed centre-of-mass
%energy and the narrow Z resonance. 
%\begin{eqnarray}  
%\Gamma^{\alpha \beta \mu}_{Z\gamma V} (q_1, q_2, P)
%&=& \frac{ i (P^2-m_V^2)}{\mzd}
%\Bigg \{ h_1^V (q_2^\mu g^{\alpha \beta}-q_2^\alpha g^{\mu \beta} )
%+ \frac{h_2^V}{\mzd} P^\alpha [ (Pq_2) g^{\mu \beta}- q_2^\mu P^\beta ]
%\nonumber \\
%&-& h_3^V \epsilon^{\mu \alpha \beta \rho} q_{2\rho}
%~-~\frac{h_4^V}{\mzd} P^\alpha \epsilon^{\mu \beta \rho
%\sigma}P_\rho q_{2\sigma} \Bigg \}~ . \label{hZgamma}
%\end{eqnarray}

%In the following, after a brief description of the OPAL detector and of the
% Monte Carlo samples used in the analysis in section
%\ref{sec:detMC}, 
%the selection of \qqbarg events and of events with missing energy 
%accompanied by an 
%isolated photon is described in section \ref{sec:selection}. 
%A measurement of the cross-section 
%for the \qqbarg and \nunug channels, within a specific 
%kinematic signal definition, is presented in 
%section \ref{sec:xsec}. 
%The analysis method used in the interpretation of the data, 
%the  resulting constraints on 
%Z$\gamma$Z and Z$\gamma\gamma$ couplings
%and the related systematic errors
%are discussed in section \ref{sec:interpretation}. 
%%Section \ref{sec:bounds} describes 
%% the results obtained with the data
%%collected at $\sqrt{s}=189$~\GeV\ 
%%uncertainties. 

%-----------------------------------------------------------------------
\section{Detector and Monte Carlo Simulation}      \label{sec:detMC}
%-----------------------------------------------------------------------
The OPAL detector, described in detail in \cite{bib:OPAL-detector},
consists of a central tracking system inside a solenoid providing a magnetic
field of 0.435 T, and of an electromagnetic calorimeter, complemented by a 
presampling system and  an array of scintillation counters for 
time-of-flight measurements; hadron calorimetry 
is obtained by instrumenting the 
magnet return yoke which is surrounded by muon chambers. 
A system of forward calorimeters extends the angular coverage of the 
detector down to a polar angle\footnote{In the OPAL coordinate system, 
$\theta$ is the polar angle defined with respect to the electron 
beam direction and $\phi$ is the azimuthal angle.}  of 24 mrad. 
However, due to the installation in 1996 
of a thick tungsten shield designed to protect the tracking chambers from 
synchrotron radiation background, the effective limit of 
electromagnetic hermeticity is around 33 mrad.
The integrated luminosity of the data samples is determined
from the rate of small angle Bhabha 
scattering events observed in the 
silicon-tungsten calorimeter \cite{bib:OPAL-SW} with a precision of 0.22\%.

Track reconstruction is performed by combining the information from a 
silicon microvertex detector, a vertex drift chamber,  a 
large volume jet drift chamber and an outer layer of drift chambers
for the measurement of the $z$ coordinate. 
The most relevant subdetector for the event topologies used in the
analysis presented here is the electromagnetic calorimeter. 
It consists of an array of 9440 
lead-glass blocks in the barrel ($|\cos{\theta}| < 0.82$) arranged in an
almost-pointing geometry and two dome-shaped end caps, each of 1132
longitudinally aligned lead-glass blocks, covering the polar 
angle range $0.81 < |\cos{\theta}| < 0.984$.
%A wide set of trigger signals\cite{bib:OPAL-TR}, overlapping in acceptance,
%based on energy deposits in the lead-glass blocks guarantees full trigger 
Trigger signals~\cite{bib:OPAL-TR},
based on energy deposits in the lead-glass blocks and also  
on a coincidence of energy in the barrel electromagnetic calorimeter
and a hit in the time-of-flight system,  guarantee full trigger 
efficiency for both the \qqbarg\ and the 
 \nunug events falling within the signal 
 definition criteria used in this analysis.

The Standard Model processes leading to the \qqbarg and 
\nunu$\gamma$ final states have been simulated using the Monte Carlo 
generators 
%\PYTHIA~\cite{bib:Pythia} and 
\kk~\cite{bib:KK} for the 
hadronic channel and \KORALZ~\cite{bib:KORALZ} and 
\NUNUGPV~\cite{bib:NUNUGPV98} for the missing energy channel. 
%%%%%%IFTN
%The \kk\ Monte Carlo provides an accurate treatment of the ISR process 
%based on the exact matrix element for one real photon radiation and for two
%or three ISR photons in the collinear approximation. Moreover, the exact
%matrix element of order $\alpha$ for initial state QED corrections and the 
%leading logarithms up to order three are exponentiated.   
%%%%%ENDIF
For the \qqbarg channel, the fragmentation, which includes photon 
radiation from the quarks, and the hadronization are simulated 
with the \Jetset\ package \cite{bib:JETSET} 
%the adjustable parameters of the model have been set according to the OPAL
%specific tuning \cite{bib:JETSETtuning} obtained from extensive studies 
%of hadronic events at the Z resonance. 
tuned on the basis of extensive studies 
of hadronic events at the Z resonance as described in~\cite{bib:JETSETtuning}.
The \grcff~\cite{bib:grc4f} generator has been used to estimate the 
background to the \qqbarg channel from four-fermion production. 
The contribution to the background from two-photon interactions has been 
studied with the Monte Carlo generators \PHOJET~\cite{bib:PHOJET}, for the 
untagged and double-tagged events, and
\HERWIG~\cite{bib:HERWIG}, for the single-tagged events and
charged current deep inelastic scattering events.   
In the \nunug channel, the contamination from  Bhabha 
events has been estimated using the \BHWIDE~\cite{bib:bhwide} 
and T{\sc eegg}~\cite{bib:teegg} generators.
The contamination from four-fermion production
has been studied using \grcff~\cite{bib:grc4f} and \KORALW~\cite{bib:koralw},
while to determine the background from two-photon interactions 
the V{\sc ermaseren}~\cite{bib:vermaseren} Monte Carlo generator has been used.
The R{\sc adcor}~\cite{bib:radcor} Monte Carlo
has been used to study  the $\rm{e^{+}e^{-} \rightarrow \gamma\gamma}$ 
background and the energy response of the calorimeter to photons.
All the Monte Carlo samples described above were processed through the
OPAL detector simulation~\cite{GOPAL}.
For the interpretation of the data, as
will be described in more detail in section 
\ref{sec:interpretation}, a Monte Carlo generator~\cite{bib:Baur},
 based on the matrix element for 
${\rm f\bar{f}\gamma}$ production in \epem\ collisions and including 
the contributions from NTGC, has been used.

%The only SM contribution missing, in the neutrino channel, is the t-channel 
%W exchange diagram which has been calculated with \NUNUGPV. 
%The effect of higher order QED corrections has been incorporated
%into the calculation using the collinear radiator function from 
%\Excalibur~\cite{bib:Excalibur}. 
%Moreover, the Monte Carlo for anomalous couplings does not implement higher 
%order QED corrections on the Standard Model processes (which are themselves
%order $\alpha$) nor on the anomalous process. 

%-----------------------------------------------------------------------
\section{Event Selection and Cross-Section Measurements} 
\label{sec:selection}
%-----------------------------------------------------------------------

\subsection{The selection of {\boldmath \qqbarg} events}
The selection of hadronic events with isolated high energy photons is 
performed on events preselected as high multiplicity hadronic 
events in a data sample corresponding to an integrated luminosity of 
176.2 ${\rm pb^{-1}}$ with an average centre-of-mass energy of 188.6~\GeV.  
The preselection criteria, described in~\cite{bib:ff172}, 
are based on the track and cluster multiplicity, 
on the visible energy and on the longitudinal imbalance of the energy measured 
in the electromagnetic calorimeter.  
%\begin{itemize}
%\item[-] total visible energy larger than 14\% of the centre of mass energy;
%\item[-] multiplicity of charged tracks higher than 4;
%\item[-] multiplicity of electromagnetic clusters higher than 6;
%\item[-] energy imbalance\footnote{This quantity is defined in terms of the
%    energy deposits measured in the electromagnetic calorimeter.} 
%    in the event, $|\sum E_i\cos\theta_i/\sum
%    E_i|$, lower than 0.75.    
%\end{itemize}
The events satisfying the preselection requirements
%, 18948, 
are processed by a photon search algorithm. 
%which selects 
%3205 of them as events containing at least one isolated 
%photon candidate.  

\begin{figure}
%\centerline{\Huge \bf OPAL preliminary}
\centering
  \epsfxsize=\textwidth
  \epsfbox[60 520 540 780]{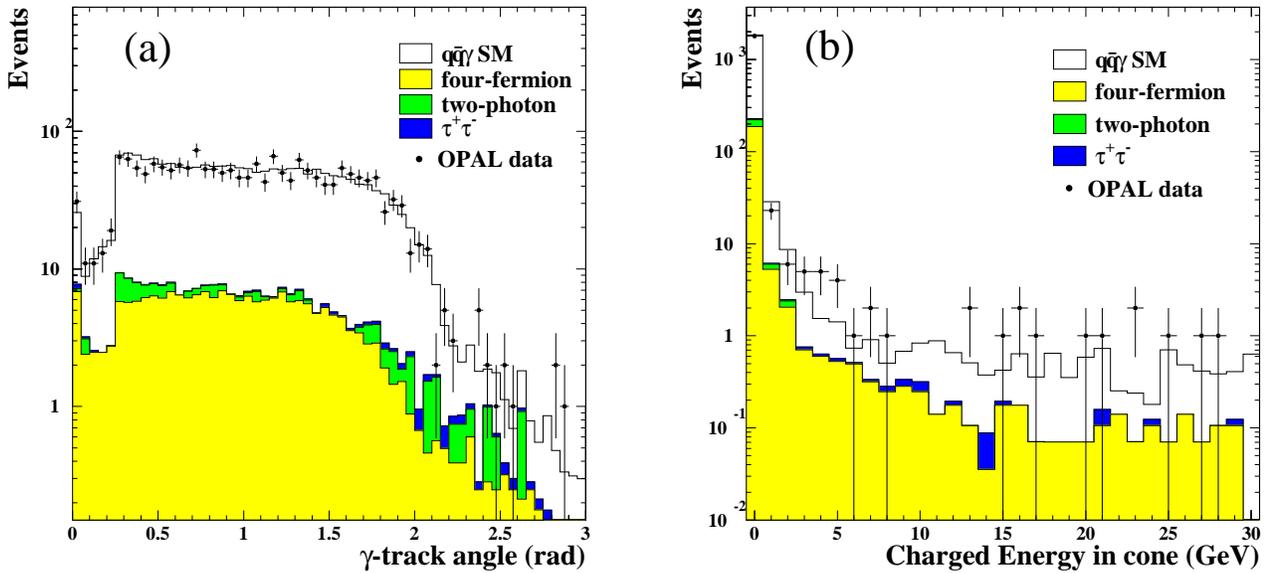}
\caption{(a) Distribution of the angle between the photon candidate 
and the closest track; the sharp edge at 0.26~rad reflects the size of the 
isolation cone. (b) Distribution of the total charged energy in the 
isolation cone; the explicit cut on this variable is set to 2~GeV. 
In both cases, the distribution refers to \qqbarg\ events which are 
 accepted or which fail only a single selection 
criterion. The Monte Carlo prediction is normalised to 
 the integrated luminosity of the data.}
\label{fig:idhega}
\end{figure}

The photon identification is based on an algorithm 
optimised for photon search in hadronic events described in~\cite{hgg172}.  
Electromagnetic clusters without associated tracks in the 
central detector are accepted as photon candidates 
if their energy is higher than 5\% of the beam energy and 
their polar angle lies in the acceptance region of the 
lead-glass calorimeter.
%, $12.5^\circ\le \theta\le 167.5^\circ$. 
The number of lead-glass blocks involved and the energy sharing 
among them are required to correspond to typical patterns 
defined for photon identification in the OPAL calorimeter. 
An isolation criterion is then applied in order to reject electromagnetic 
clusters associated with jets. The total energy deposition 
in the electromagnetic calorimeter (not associated to the photon candidate) 
within a cone of $15^\circ$ around the photon 
flight direction is required to be less than 2~\GeV.
In addition, 
the sum of the momenta of tracks which, extrapolated to 
the calorimeter surface, 
fall inside a  $15^\circ$ cone around the photon impact point is also 
required to be lower than 2~\GeV. 
Finally, systems of one or two well reconstructed tracks associated with
electromagnetic clusters,  consistent with a photon conversion
according to the criteria described in~\cite{hgg183}, are 
included if the reconstructed photon satisfies the criteria listed above. 
%%%%%%%%%%%%%%%%%%%%Photon search algo quality statement%%%%%%%%%%%%%
%The photon search algorithm has been tested on a sample of 
%\epem$\rightarrow {\rm  Z/\gamma^\star}\rightarrow$\qqbar\ Monte Carlo 
%events, and has been found to identify genuine ISR photons with 
%a purity of 90\%. The contamination, arising from photons 
%related to the final state 
%(either true final state radiation or photons from decays of particles,
%especially $\pi^0$, in the jets)
%or ``fake photons'',  is concentrated at energies lower than 20~\GeV. 
%%%%%%%%%%%%%%%%%%%%%%%%%%%%%%%%%%%%%%%%%%%%%%%%%%%%%%%%%%%%%%%%%%%%%%
The photon identification algorithm has been extensively 
studied in order to assess the level of accuracy of the 
estimate of the efficiency obtained from Monte Carlo 
\qqbarg\ events. 
In particular, the selection criteria of the 
algorithm have been adjusted 
in order to minimise their sensitivity to unsatisfactory 
modelling in the simulation. 
A residual discrepancy in the identification 
efficiency for converted photons has been observed and 
taken into account as a correction of 1.25\% to the overall 
selection efficiency.
As an example of the quality of the modelling of 
the photon identification algorithm, 
figure \ref{fig:idhega} shows
the angle between the photon candidate and the closest 
track in the event and the total charged energy in the isolation cone. 

After the photon search, all the clusters and tracks in the event which
are not associated to the most energetic photon candidate are grouped 
into jets according to the Durham $k_{T}$ \cite{bib:Durham} scheme with 
resolution  parameter $y=0.02$.
If more than four jets are reconstructed, the event is forced to 
have four jets in addition to the isolated high energy photon. 

%Since the event topology for the anomalous process  
%consists of one isolated photon recoiling against a system of 
%$q^2={\rm M_Z}^2$, the contamination of
%the low energy tail is dangerous in the case of events with a very
%energetic photon and one or more photon candidates of lower energies.
%This class of events represents ?\% of the sample selected in the
%data, and since the overall resolution in the reconstruction of hadronic 
%events and the kinematical constraints are not stringent enough to easily 
%disentangle a pure sub-sample (from Standard Model Monte Carlo studies the expected 
%fraction in which the soft photons are not ISR is of the order of ??? 50\%),  
%it is not used in the analysis. 

In \epem$\rightarrow \rm Z\gamma$ at
  $\sqrt{s}=189$~GeV the photon energy spectrum is peaked at 
  approximately 72~GeV,
  reflecting the sharp Z resonance.
  In order to select the topology corresponding to a high energy 
  photon recoiling against a hadronic system of invariant mass equal 
  to the Z boson mass, the signal definition is based on kinematic 
  cuts applied to the most energetic photon in the event:
\begin{itemize}
\item[~] 50~\GeV\ $\rm{<E_\gamma<}$ 90~\GeV\ ; 
\item[~] 15$^\circ$ $<\theta_\gamma<$ 165$^\circ$,
\end{itemize}
where $\rm{E_\gamma}$ and $\theta_\gamma$ are the photon energy and polar 
angle, respectively.
%The acceptance cuts are defined aiming to optimal event reconstruction and 
%enhancement of the sensitivity to NTGC.
%signal from AGC with respect to the dominant SM 
%background from radiative return to the Z$^0$ resonance. 

To improve the photon energy resolution and 
 suppress further the background 
from non-\qqbarg events and the feedthrough from events outside 
the signal definition, a kinematic fit is applied to all the events with 
at least one photon of energy larger than 30~GeV. 
The fit~\cite{bib:ff172} imposes energy and momentum conservation using 
as input the photon  and the jet momenta. 
Undetected ISR is allowed to 
compensate for missing longitudinal momentum in the beam pipe region
if the $\chi^2$ probability of the fit is smaller than 1\%. 
 
From a study of Monte Carlo \qqbarg events, 
the kinematic fit improves the photon energy resolution 
by a factor two.
The events for which the fit converges (99.5\%) are finally selected if 
the fitted values of $E_\gamma$ and  $\theta_\gamma$
 satisfy the signal definition cuts. 
In order to suppress contamination from photons originating from the
jets, only the sub-sample in which  
$\alpha_{\gamma-{\rm jet}} > 30^\circ$, where $\alpha_{\gamma-{\rm jet}}$
is the angle between the photon and the closest jet, is retained for the 
analysis.

The selection efficiency and the feedthrough in the kinematic acceptance 
are listed in table~\ref{tab:eff}. They have been 
estimated from fully simulated \epem$\rightarrow$\qqbar\  
Monte Carlo events, where the signal-like topology arises 
from radiative return to the Z resonance. 
The efficiency has been calculated with respect to the
kinematic signal acceptance defined above. 
The feedthrough represents the fraction of selected events which do not
belong to the kinematic acceptance; it is mainly due to resolution 
effects, but it includes also a small contamination (0.35\%) due 
to non-ISR photon candidates 
in \epem$\rightarrow$\qqbar\ events falling in the signal 
acceptance after reconstruction. 
Both the efficiency and the feedthrough are corrected 
for the aforementioned residual disagreement between the 
performance of the photon search algorithm in the data and in the 
Monte Carlo.  

\begin{figure}
%\centerline{\Huge \bf OPAL preliminary}
%\vspace{1.cm}
  \epsfxsize=\textwidth
  \epsfbox[0 0 567 567]{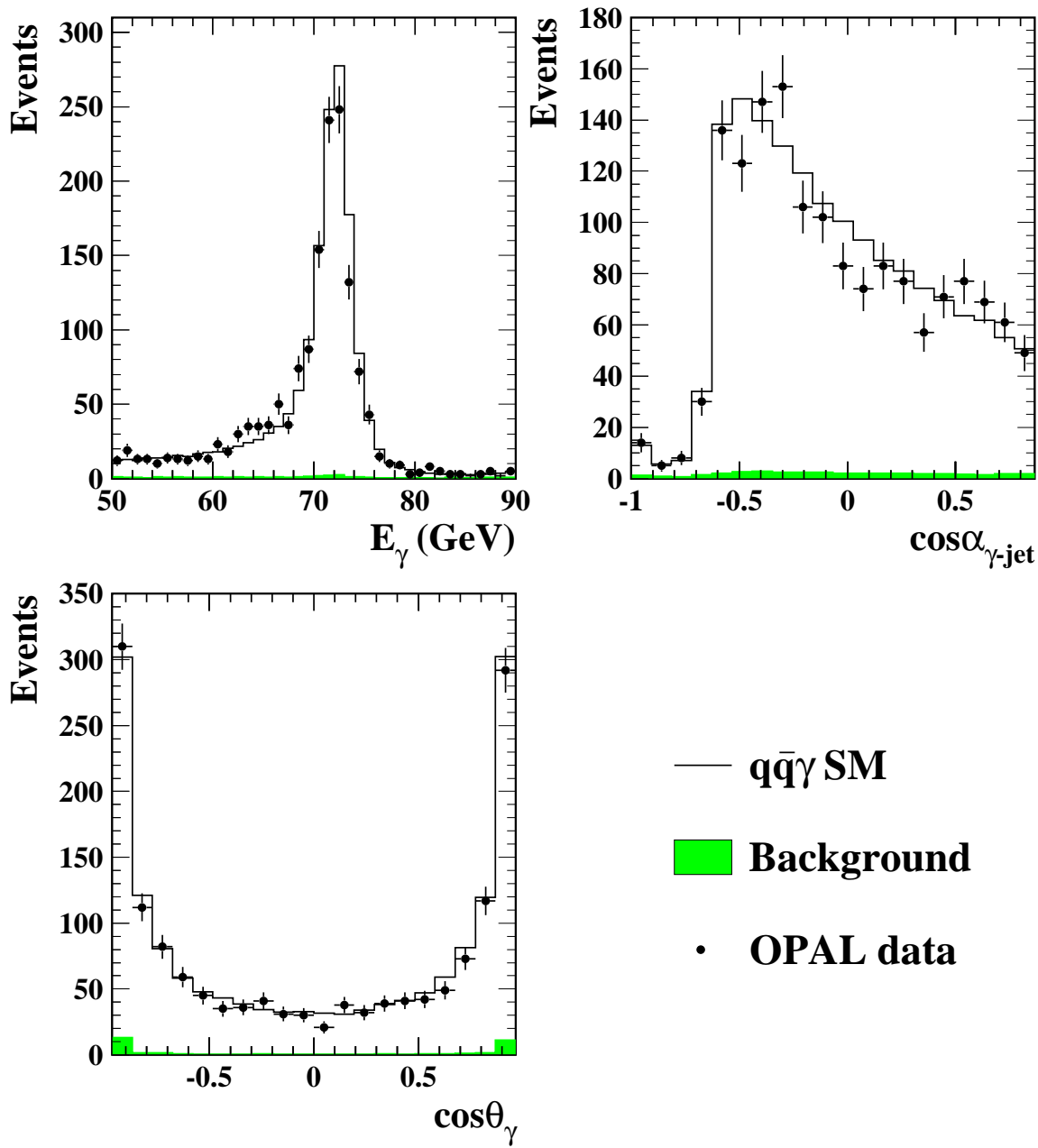}
  \caption{Distribution of 
  ${\rm E_{\gamma}}$,  $\rm \cos{\alpha_{\gamma-jet}}$  
  and  $\rm \cos\theta_\gamma$  measured in 
  \qqbarg events. 
  The data (dots) are superimposed on the Standard Model 
  predictions from the \kk\ Monte Carlo (solid histogram),
  normalised to the integrated luminosity of the data sample. The shaded
  area in the histograms represents the background from four-fermion
  production and two-photon interactions and the contamination from photon
  candidates not corresponding to truly initial state radiation in 
  \epem$\rightarrow$\qqbar\ events.}
\label{fig:smdata}
\end{figure}

The numbers of events selected in the data and in the Monte Carlo samples 
are listed in table~\ref{tab:events}. Figure \ref{fig:smdata} shows 
the distributions
of the photon energy and polar angle and angular separation with respect to
the closest jet for the events selected in the data, 
compared with the Standard Model
expectation from Monte Carlo.
The total background (2.5\% of the selected events) comes from 
four-fermion production (1.59\%)  two-photon interactions (0.59\%) and
tau-pair production (0.34\%). 
%~\foortnote{``Non radiative'' \qqbar\  are defined 
%in the Monte Carlo as having a true $s'/s=(1-\frac{2E_{\gamma}/E_{ecm}})
%\gt 0.83$} (0.35\%) in which 
%photons associated to the final state are identified as ISR photons.  
The agreement between the data and the Monte Carlo is 
in general satisfactory, except perhaps in the photon energy distribution  
where a slight deficit of events is observed 
in the radiative return peak.
% Extensive studies 
%on the quality of the photon reconstruction 
%have been made, but no evidence of systematic
%effects, which could be responsible for such discrepancies,
%has been found.

\subsection{The selection of {\boldmath \nunug} events}

The selection of events with an isolated high energy photon 
accompanied by missing energy and low activity in the detector 
follows the single-photon analysis 
described in \cite{bib:nng189}. 
The data sample used in the analysis 
corresponds to an integrated luminosity of 
177.3 ${\rm pb^{-1}}$, with an average 
centre-of-mass energy of 188.6~\GeV.  
After the single photon selection, 643 events are retained in the data. 
The same additional conditions
which define the kinematic acceptance 
in the $\rm{q \bar q \gamma}$ channel,
$50~\rm{GeV}<E_\gamma<90~\rm{GeV}$ and $15^\circ<\theta_\gamma<165^\circ$,
are then applied to the most energetic photon in the electromagnetic 
calorimeter.
%Briefly,  photons are identified as clusters of energy deposited 
%in the lead-glass electromagnetic 
%calorimeter, while the tracking system is used to reject events 
%containing prompt charged particles. 
%Full hermeticity in the electromagnetic calorimeter coverage
%is achieved in the forward region with the use of the
%the gamma-catcher calorimeter, the forward 
%calorimeter and the silicon-tungsten calorimeter, which 
%p%rovide powerful rejection against the otherwise 
%dominant radiative Bhabha background.
%Scintillators in the barrel (Time of Flight) and endcap (Tile Endcap) 
%regions are used to efficiently reject backgrounds from cosmic ray 
%interactions, by providing time measurements for 
%the large fraction of
%photons which convert in the material in front of the 
%calorimeter. Beam related backgrounds 
%are rejected using again the scintillator timing 
%measurements and also the information 
%from the electromagnetic calorimeter
%shower shape, the hadron calorimeter and the muon detectors. 

The predicted efficiency and  the feedthrough
in the kinematic acceptance are listed in 
table~\ref{tab:eff}. They have been estimated averaging the 
predictions of the \KORALZ\ and the \NUNUGPV\ Monte Carlo,
which in the signal acceptance agree 
within (0.2$\pm$0.5)\% in the efficiency
and  (1$\pm$9)\% in the feedthrough.
%The kinematic cross-sections also agree within a relative 
%(1.6$\pm$1.1)\%, where the error comes from Monte Carlo statistics. 

\begin{figure}
%\centerline{\Huge \bf OPAL preliminary}
%\vspace{-0.5cm}
  \epsfxsize=\textwidth
  \epsfbox[0 230 567 567]{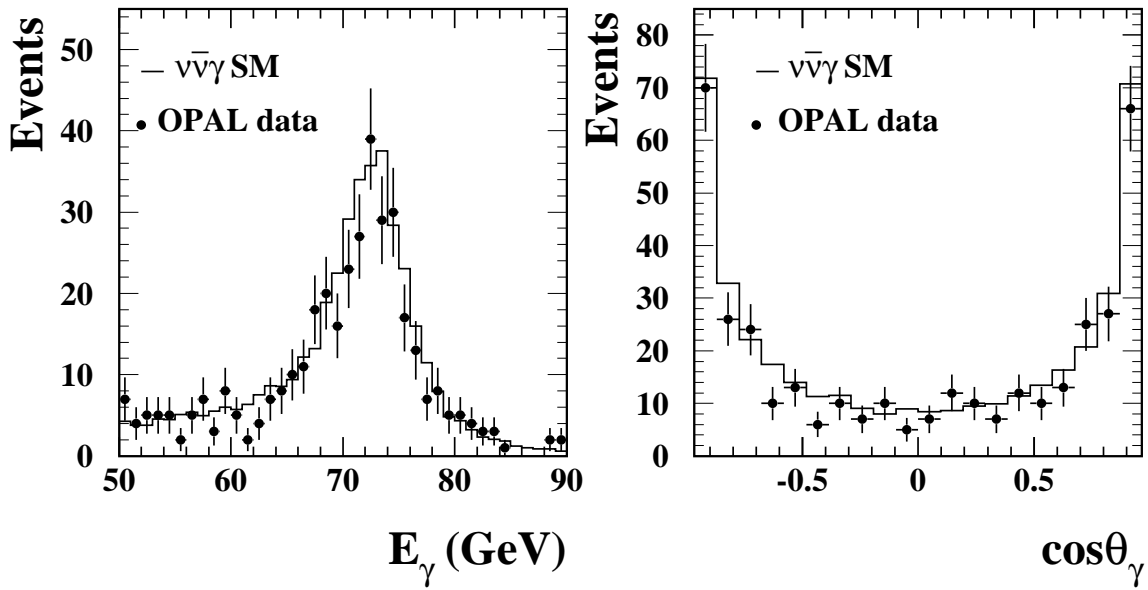}
  \caption{Distribution of kinematic variables 
  ${\rm E_{\gamma}}$ and $\rm \cos\theta_\gamma$ measured in  \nunug events. 
  The data (dots) are superimposed on 
  the Standard Model prediction (solid histogram), 
  taken as the average of the two generators \KORALZ\ and \NUNUGPV,
  normalised to the integrated luminosity of the data sample.
  The negligible ($<$0.3\%) background from physics processes other than \nunug
  is not shown.}
\label{fig:smdatanng}
\end{figure}

The number of events selected in the data 
are listed in table \ref{tab:eventsnng}, together with 
those expected from all the relevant physics processes and 
from instrumental backgrounds. 
As potential sources of physics background, 
four-fermion processes, radiative Bhabha and 
$\rm{e^+e^- \rightarrow \gamma\gamma}$ events
have been considered, while the residual 
cosmic ray and beam-related contamination
have been  estimated using control samples enriched in these 
backgrounds.   
Both these sources result in an overall negligible (0.24\%) contribution 
to the selected events. 
Figure~\ref{fig:smdatanng} shows the distributions
of the energy and of the polar angle of the photon for 
the selected events, compared with the Standard Model
expectation.

%-----------------------------------------------------------------------
\subsection{Cross-section Measurements} \label{sec:results}
%-----------------------------------------------------------------------
\label{sec:xsec}
From the number of observed events in each channel 
and from the predicted 
efficiency, feedthrough and backgrounds, 
 as presented  in
sections~\ref{sec:selection}.1 and \ref{sec:selection}.2, 
the cross-sections within the kinematic signal acceptance 
are  measured to be:
\begin{eqnarray*}
 \rm  \sigma_{q\bar{q}\gamma} & = &9.42 \pm 0.25~\stat \pm 0.15~\syst\ \:\:
 {\rm pb} \\
 \rm  \sigma_{\nu\bar{\nu}\gamma} &= &2.52 \pm 0.13~\stat \pm 0.05~\syst\ \:\:
 {\rm pb}.
\label{eqn:xsec}
\end{eqnarray*}
These measurements are in reasonable agreement, within the errors which are 
dominated by the statistical
uncertainty, with the predictions 
from the Standard Model, which are respectively   
$\rm  \sigma_{q\bar{q}\gamma}^{SM} = 9.75 \pm 0.03$~pb
and $\rm  \sigma_{\nu\bar{\nu}\gamma}^{SM} = 2.81 \pm 0.02$~pb. 
The predicted \qqbarg\ cross-section is obtained from \kk, while the \nunug\
cross-section is the average of the predictions of \KORALZ\ and \NUNUGPV\
which are consistent within $(1.6\pm1.1)\%$. The errors 
associated with the Standard Model predictions come from the 
Monte Carlo statistics.
Taking into account the experimental systematic uncertainties 
discussed in the following section, the overall discrepancy corresponds to 
$-1.1$  standard deviations for the 
\qqbarg  channel and $-2.1$ for the \nunug channel.

\subsection{Systematic Errors}
\label{sec:sysxs}
The different sources of systematic uncertainties 
affecting the cross-section  
measurements are summarised in table~\ref{tab:sysxs} and 
are discussed in the following:
\par
\bigskip
\noindent
Systematics specific to the \qqbarg channel:
\begin{itemize}
\item[-] The uncertainty on the selection efficiency.
  Contributions to this uncertainty come from imperfect 
  modelling of the material in the detector, affecting 
  the estimate of the photon conversion rate (0.77\%), 
  from inaccuracies in the modelling of the identification 
  efficiency for converted photons (0.62\%),   
  from uncertainty in the simulation of the track-cluster 
  association at forward angles (0.64\%) and 
  from the photon isolation criterion (0.26\%). 
  The total uncertainty coming from the modelling of the 
  selection efficiency translates into a relative error 
  on the cross-section of 1.2\%. 
  The uncertainties arising from limited Monte Carlo statistics 
  used to evaluate the efficiency  
  and the feedthrough are respectively 0.14\% and 0.07\%.
\item[-] The modelling of the jet reconstruction.
  An additional smearing of the jet
  energies and directions and a shift of the jet energy scale
  has been  applied in the Monte Carlo, 
  on the basis of an extensive comparison
  of two-jets events in calibration data collected 
  at the $\rm Z^0$ peak and in the simulation.   
  The resulting variation 
  (0.36\%) in the cross-section has been assigned as a systematic error.
\item[-]   The sensitivity of the analysis to the jet multiplicity in the 
  event. The results of a modified analysis, where each event has been 
  forced to contain exactly two jets in addition to the isolated photon,
  have been compared with those from the standard analysis. 
  The difference has been assigned as a systematic error (0.69\%).
\item[-]   The uncertainty related to the models used to simulate 
  the hadronisation process. This uncertainty (0.08\%) has been  evaluated 
  comparing the results  when either  the \Jetset\ or the \Herwig\ 
  hadronisation schemes have been used in the Monte Carlo.
\item[-] The uncertainty (0.61\%) due to the background subtraction; this
  is dominated by the 100\% uncertainty on the normalisation of the background
  from two-photon interactions  as predicted by \Herwig, which
  is expected to give the best description of two-photon interactions in
  the data, and by \Pythia.
  This large uncertainty is assigned to
  cover possible mismodelling of the hard-fragmentation processes
  in the very small fraction of the two-photon cross-section 
  retained in the selection, 
  as suggested by a comparison with 
  the {\sc F2GEN}~\cite{bib:F2GEN} generator. Finally, 
  the  uncertainty related to the contamination from fake ISR   
  photons in non-radiative \qqbar\ events
  is estimated to be 0.35\%.
\end{itemize} 

Systematics specific to the \nunug channel:
\begin{itemize}
\item[-]
The uncertainty on the selection efficiency.
Systematic errors come 
from the estimate of the detector occupancy (1\%) and from the 
imperfect description of the conversion probability and 
of the conversion tagging efficiency in the Monte Carlo (0.8\%).
A 0.9\% uncertainty is assigned to account for 
systematic effects in the simulation of the cuts used to reject 
radiative Bhabha and  $\rm{e^+e^- \rightarrow \gamma\gamma}$ 
events.
The uncertainty on the efficiency of the timing cuts 
used to reject background from cosmic rays translates into 
a systematic error of 0.5\%, while  
an uncertainty of 0.6\% is assigned 
to cover possible mismodelling of the other cuts complementing  
the timing requirements in the cosmic and 
instrumental background rejection.   
The overall uncertainty on the modelling of the selection efficiency 
is estimated to be 1.7\%.
Finally, the statistical uncertainty on the efficiency and feedthrough 
as estimated from Monte Carlo contributes with a 
0.3\% uncertainty on the cross-section.
\item[-] The uncertainty on the background from  
 processes other than \nunug, amounting to 0.25\%.
 This includes a 50\% systematic uncertainty on the expected background, 
 which was assigned to cover possible mismodelling of the vetoes 
 used in the rejection.
\end{itemize}

Systematic uncertainties common to both analyses:
\begin{itemize} 
\item[-] The uncertainty on the 
   angular acceptance arising from  residual biasses in the 
   coordinate reconstruction and 
   from the absolute knowledge of the detector geometry.
   The overall uncertainty on the 
   position of the electromagnetic showers, 
   estimated in~\cite{bib:2f189} to be 0.001~rad,
   results in a relative uncertainty of 0.19\%. 
\item[-] 
  The effects of mismodelling of the energy scale and the
  resolution of the electromagnetic calorimeter. They have been investigated 
  using a control sample of about one thousand strongly collinear 
  $\rm{e^+e^- \rightarrow \gamma \gamma}$ events, whose energies
  are expected to be very close to the beam energy. 
  The Monte Carlo has been found to reproduce the electromagnetic calorimeter 
  energy scale 
  and resolution in the data 
  within respectively 0.3\% and 10\%.
  These differences are in very good agreement 
  with those observed using as a reference process a sample of
  wide angle Bhabha events~\cite{bib:2f189}.
  The systematic error has then
  been estimated by modifying the absolute energy scale and  
  resolution in the Monte Carlo within these limits 
  and the corresponding variation of the cross-section 
  has been taken as systematic error. 
 
\item[-] The uncertainty on the beam energy ($\pm20$~MeV).
  The effect on the cross-section was evaluated by 
  appropriately scaling the energy of the most energetic photon
  in the Monte Carlo, while
  leaving unchanged the invariant mass of the system 
  recoiling against the photon.   

\item[-] The uncertainty on the measurement of the integrated luminosity,
  0.22\%.

\end{itemize}

\section{Data Interpretation}
\label{sec:interpretation}
Since the data presented in the previous section show no evidence 
of deviations with respect to the Standard Model expectation,
they are used to derive
bounds on the strength of anomalous $\rm Z\gamma Z$ and $Z\gamma\gamma$ 
couplings. 

\subsection{Analysis Procedure}

The analysis is based on a comparison of the measured event rate and of 
the differential distributions 
with the theoretical predictions for 
the Standard Model processes and possible contributions from NTGC.
The energy spectrum and the cosine of the 
polar angle of the photon are used in both channels, while 
in the \qqbarg channel the cosine of the angle 
between the photon and the closest jet is also used.
Due to the almost monochromatic photon energy spectrum,
%In the basic three-body kinematics of such events, 
this angle is 
strongly correlated to the quark emission angle in 
the Z decay rest frame, which is sensitive to NTGC.

The theoretical predictions as a function of different values of the 
anomalous  couplings $\rm h_{\it i}^{Z,\gamma}$ are obtained from a Monte Carlo 
generator~\cite{bib:Baur} for ${\rm f\bar{f}\gamma}$ production in 
\epem\ collisions. This generator is based on the full matrix element, 
in the lowest order approximation, 
for all the relevant Standard Model processes and for processes 
generated by anomalous
trilinear neutral gauge couplings.  
The only Standard Model contribution missing in the calculation 
is the t-channel W boson exchange in the \nunug channel, which 
has been estimated from \NUNUGPV\ and included
in the reweighting procedure described in the following. 
The effect of higher order QED corrections from initial state radiation,
which have been found to reduce the contribution from anomalous couplings 
by typically 15\%, has been 
incorporated into the calculation using a collinear radiator function 
from the \Excalibur~\cite{bib:Excalibur} Monte Carlo.
 
%To check the quality of the ISR description, the predictions of 
%the improved calculation for $\rm h_{\it i}=0$ have been compared 
%to the reference Standard Model calculations used in the two analyses,
%which include a more refined treatment of the QED corrections.  
%In both the \qqbarg and \nunug channels, an 
%agreement better than 1\% in the cross-sections 
%has been observed within the kinematic acceptance 
%adopted in the signal definition.
%For \nunug, such test has been performed 
%taking as a reference the \nunumg~channel, to disentangle the 
%discrepancy arising from the missing W contribution in the 
%\nunueg~channel from those arising from the quality of the ISR 
%treatment. 

%Finally, the theoretical predictions has been 
%corrected for detector acceptance and resolution effects. 
%The corrections have been obtained from Monte Carlo \qqbar\ and
%$\nu\bar{\nu}$ events simulated with the generators \kk, \NUNUGPV\ and
%\KORALZ\ passed through a full simulation of the OPAL detector
%and processed with the same reconstruction procedure adopted for the data 
%analysis. 
%The corrections have been obtained from the fully simulated \qqbarg and
%\nunug  Monte Carlo events used for the cross-section measurement.
%Some channel-dependent details of this smearing procedure 
%are discussed in the following.  \newline
%\label{sec:smearing}
%The comparison of the results obtained by applying
%the signal definition cuts at the generator level and after the event 
%reconstruction, which is identical to the real data reconstruction
%procedure. 

\begin{figure}
%\centerline{\Huge \bf OPAL preliminary}
%\vspace{1.cm}
  \epsfxsize=\textwidth
  \epsfbox[0 0 567 567]{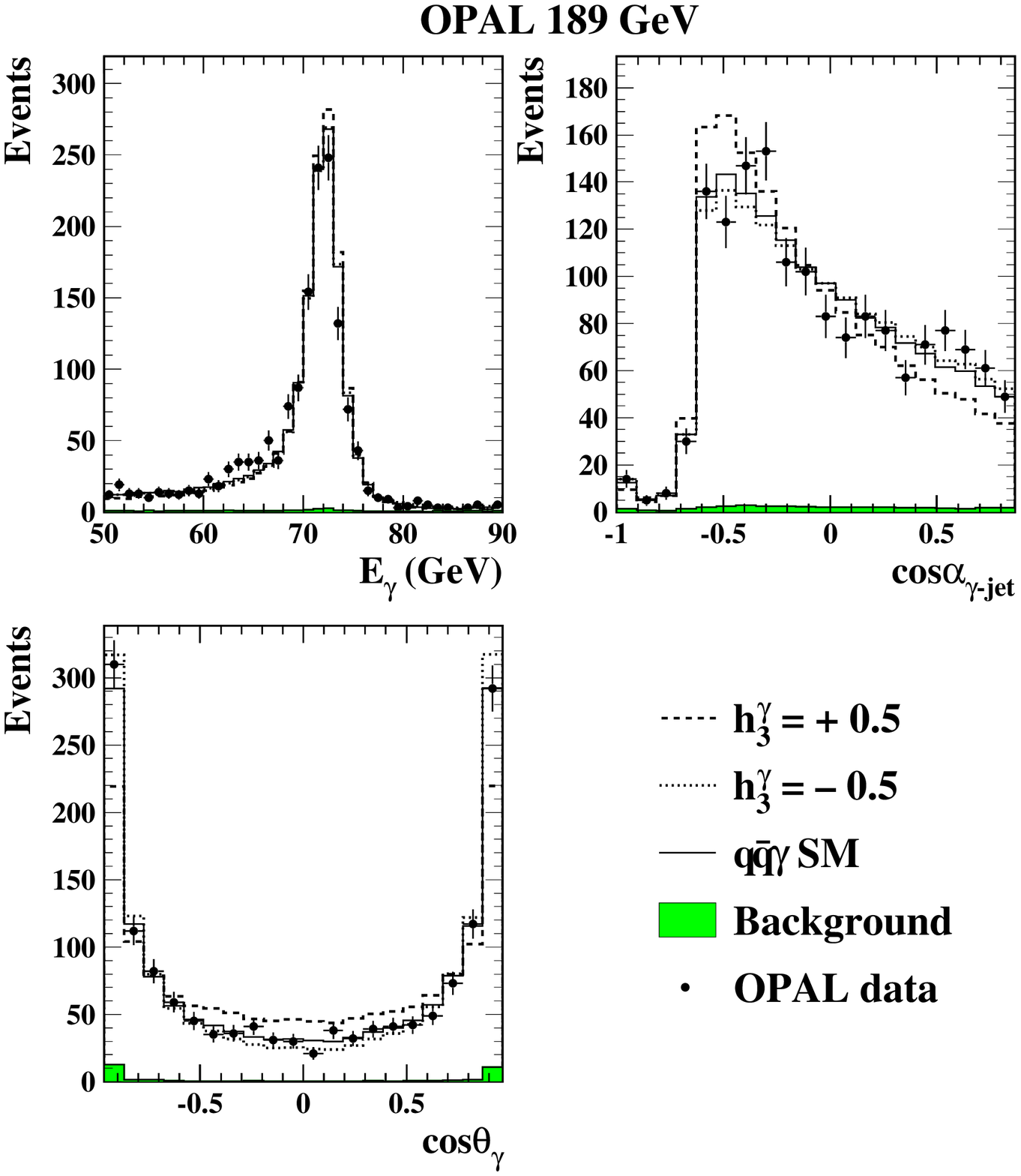}
  \caption
{Distribution of kinematic variables 
  ${\rm E_{\gamma}}$,  $\rm \cos{\alpha_{\gamma-jet}}$  
   and  $\rm \cos\theta_\gamma$ measured in 
  \qqbarg events in the data (dots) and in the \kk\ Monte Carlo
  (solid histogram).
  The simulated Standard Model expectations, reweighted   
  to incorporate the contributions from 
  $\rm {h_3^{\gamma}}=\pm 0.5$  
  are also shown. 
  The normalisation of the Monte Carlo predictions is defined by the number
  of events selected in the data.
}
\label{fig:agcdataqqg}
\end{figure}

\begin{figure}
%\centerline{\Huge \bf OPAL preliminary}
%\vspace{-0.5cm}
\epsfxsize=\textwidth
\epsfbox[0 230 567 567]{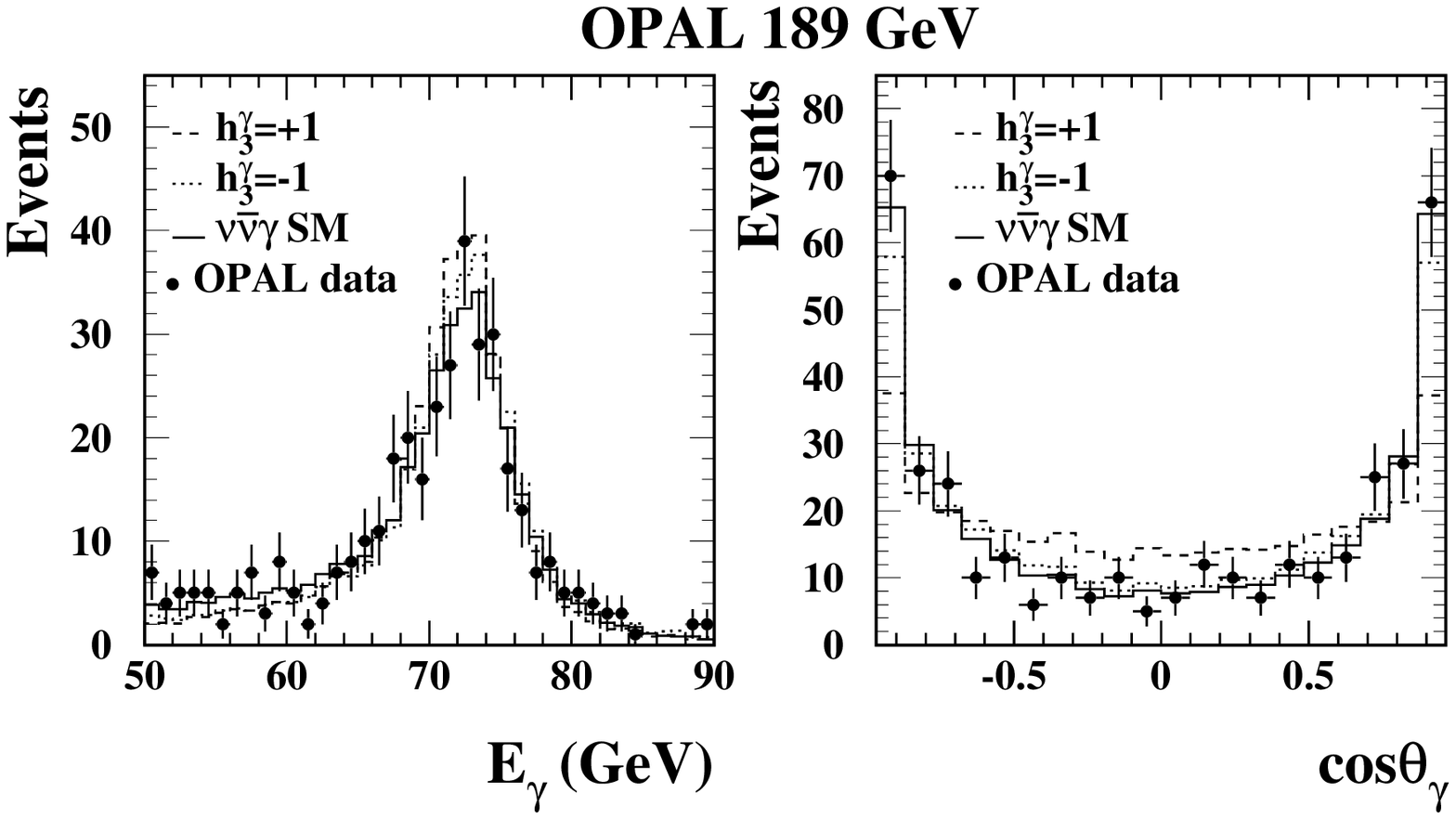}
\caption{Distribution of kinematic variables
 ${\rm E_{\gamma}}$ and $\rm \cos\theta_\gamma$ measured 
 in \nunug events in the data (dots), and in the Standard Model 
 Monte Carlo (solid line). 
 The simulated Standard Model expectations, reweighted   
 to incorporate the contributions from 
 $\rm {h_3^{\gamma}}=\pm 1$  are also shown.  The normalisation of the Monte 
 Carlo predictions is defined by the number
 of events selected in the data.}
\label{fig:agcdatanng}
\end{figure}

%\noindent
%\underline{\bf \qqbarg channel}  
The events selected in the data 
have been  classified in $5\times 4\times 4$ unequal bins of 
the three-dimensional 
$\rm (E_\gamma,\cos\theta_\gamma,\cos\alpha_{\gamma-jet})$ space.
In the case of the \nunug channel, the angle between the photon and the 
$\rm{Z}$ decay products is not experimentally accessible
% variable equivalent to 
%$\rm \alpha_{\gamma-jet}$ in the \qqbarg channel is unmeasured 
and therefore integrated out.  
%The width of the bins is not constant for each
%variable and is chosen in such a way to retain sensitivity to the effects
%of anomalous couplings on the differential distributions while keeping a 
%reasonable number of events in each cell. The average population of the 
%cells is 15 events.
The NTGC-dependent theoretical prediction for the population 
in each cell, provided by the Monte Carlo calculation, 
has been  modified to allow for the reconstruction 
efficiency and resolution effects, %asfollows:
%$$  \rm N^{\it i}_{det}(h) = \sum_{\it j} N^{\it j}_{th}(h) \times C_{\it ij} %$$
%where $\rm C_{\it ij}$ 
%represents the probability of reconstructing in the cell 
%$i$ an event belonging to the cell $j$. These correction factor with 
as determined from a large sample of fully simulated Z$^0/\gamma\rightarrow 
{\rm q\bar{q}}$ and Z$^0\rightarrow {\rm \nu\bar{\nu}}$ Standard Model 
 Monte Carlo events.  
%Feedthrough from events outside
%the signal acceptance is taken into account by defining a wider region 
%in the space of the kinematic variables.
%The probability of migration of events from the peripheral cells to the 
%internal ones, which belong to the signal definition region, is then computed
%along with the cross-migration between internal cells. 
%The available Monte Carlo statistics allows to determine the correction
%factors with a  relative precision which, for $C_{ii}$, approaches and, in
%the most populated bins exceeds, 5\%. 
%In order to minimise the impact on the analysis of uncertainties 
%inherent in the process of smearing the theoretical calculation, 
The total number of expected events and the population   
of each cell as a function of the anomalous 
couplings, $\rm N(h_{\it i})$,  is determined by reweighting 
the number $\rm{N_{SM}}$ of accepted events 
predicted by the fully simulated Standard Model Monte Carlo according to: 
\begin{equation}
{\rm N(h_{\it i}) = N_{SM}(1+\delta(h_{\it i}))}, \:\:\:\:\:\:\:\:\:\:
 {\rm \delta(h_{\it i}) = \frac{N_{rec}(h_{\it i})-N_{rec}(h_{\it i}=0)}
 {N_{rec}(h_{\it i}=0)}},
\label{eqn:expqqg}
\end{equation} 
where $\rm N_{rec}$ is the number of reconstructed events 
from the NTGC-dependent theoretical prediction, 
modified for efficiency and resolution as explained above.
Figures~\ref{fig:agcdataqqg} and \ref{fig:agcdatanng} show how the 
distributions of the kinematic variables,  folded with detector 
effects, are modified by a particular choice of 
anomalous couplings in the \qqbarg and
\nunug channels. In order to disentangle
the effect of NTGC on the event rate and on the differential distributions
the latter have been normalised to the number of events selected in the
data. 

The theoretical expectations for the event rate and the differential
distributions have been fitted to the data independently for 
the two channels, under the hypothesis that only one coupling 
at a time is non-zero. 
The most probable values of the anomalous couplings are determined 
by minimising the  negative log-likelihood defined as follows: 
$$\rm - Log L = -Log P(N^{obs},N(h)) - \sum_{\it j} Log 
P(N^{obs}_{\it j},N_{\it j}(h)) $$
where $\rm P(N^{obs},N(h))$ is the Poisson probability of observing the 
number of events $\rm N^{obs}$ if the expectation is $\rm N(h)$.  
The index $j$ runs over the number of cells defining the multidimensional 
distributions and the condition 
$\sum_j {\rm N}_j({\rm h}) = {\rm N^{obs}}$ is imposed to 
disentangle the contributions to the 
 likelihood from event rate and distributions. 

%The fit procedure has been tested on Monte Carlo samples of Standard Model 
%events reweighted for NTGC effects. The results of the fit has been found to 
%reproduce the values of
%the NTGC parameters used in the Monte Carlo samples within the statistical
%precision. Tests performed on several Standard Model Monte Carlo 
%samples of size corresponding to the data
%luminosity show that the fluctuations of the central values of the
%couplings determined by the fit are consistent with the statistical error 
%expected on the corresponding NTGC parameter in case of perfect agreement
%between data and  Standard Model predictions. 

The fit procedure has been tested on Monte Carlo samples of Standard Model 
\qqbarg and \nunug\ events reweighted for NTGC effects. 
The central values of the fit results have been found to 
correctly reproduce the values of the input NTGC parameters.
In order to check the reliability of the errors on the fit results, 
tests have been performed on several Standard Model Monte Carlo 
samples of size corresponding to the data
luminosity.  The distributions of the central values 
of the couplings determined by the fit are found to be consistent 
with those expected from the statistical sensitivity. 

\subsection{Results on Trilinear Neutral Gauge Couplings}
%\label{sec:bounds}
The values of the anomalous couplings and the statistical errors obtained
from the likelihood fit of the \qqbarg and \nunu$\gamma$ event 
rate and distributions are listed in table 
\ref{tab:agc}, together with the expected 
statistical sensitivity of the analysis which would be achieved 
in the case of perfect agreement between data and the Standard Model 
predictions. 
%For the \qqbarg channel, the results of 
%two analogous analyses, one of them based on the kinematic fit of 
%the event, the other on the direct measurements of the kinematic 
%variables, are reported in table 
%\ref{tab:agc}. The comparison shows compatible central values 
%and errors for all the  couplings in the two methods. 
In general, the total event rate  and the 
differential distributions have similar sensitivities to NTGC.
However, due to the quadratic dependence of 
the cross-section on the couplings, the fit of the event rate can
only determine the value of the couplings with a two-fold ambiguity.  
In the case of the CP violating couplings, 
$\rm h_{\it i}^{Z,\gamma}$ ($i=1,2$), 
which lead to amplitudes that do 
not interfere with the Standard Model amplitudes, the sign of the couplings is 
completely undefined, but both the cross-section and the distributions 
provide a determination of the absolute value of the couplings.
In the case of the CP conserving couplings, 
$\rm h_{\it i}^{Z,\gamma}$ ($i=3,4$), 
the interference with
the Standard Model amplitudes results in distributions 
of the kinematic variables which produce a unique minimum in the associated 
$\rm -\log L$ function, thus removing the two-fold ambiguity arising from
the event rate information. 
%the interference with the Standard Model amplitudes 
%entails different absolute
%values for the two points of minimun of the asymmetric $\rm -\log L$ 
%function based on the event rate only. 
%The distributions of the kinematic variables plays here an important role 
%due to the existence of an absolute minimum in the associated 
%$\rm -\log L$ function which, in the combination with the event rate 
%information, removes the twofold ambiguity. 

\begin{figure}
%\centerline{\Huge \bf OPAL preliminary}
%\vspace{-0.5cm}
  \epsfxsize=\textwidth
  \epsfbox[0 0 567 567]{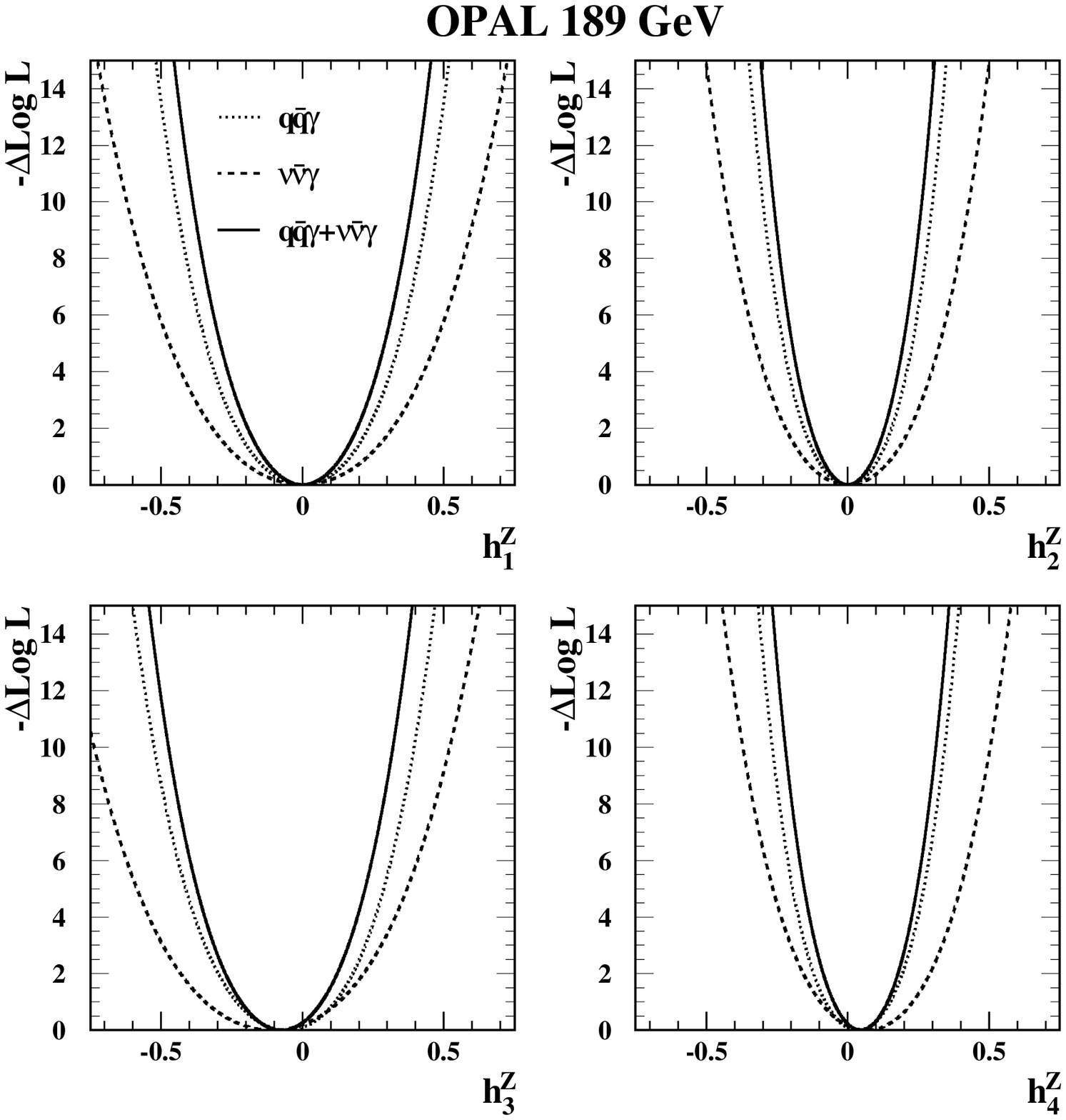}
  \caption
{Negative log-likelihood function for the ${\rm h_{\it i}^{Z}}$ couplings
  as obtained from the analysis of the \qqbarg channel 
 (dash-dotted line), of the
  $\nu\bar{\nu}\gamma$ channel (dashed line) 
  and from their combination (solid line).}
\label{fig:likz}
\end{figure}

\begin{figure}
%\centerline{\Huge \bf OPAL preliminary}
%\vspace{-0.5cm}
  \epsfxsize=\textwidth
  \epsfbox[0 0 567 567]{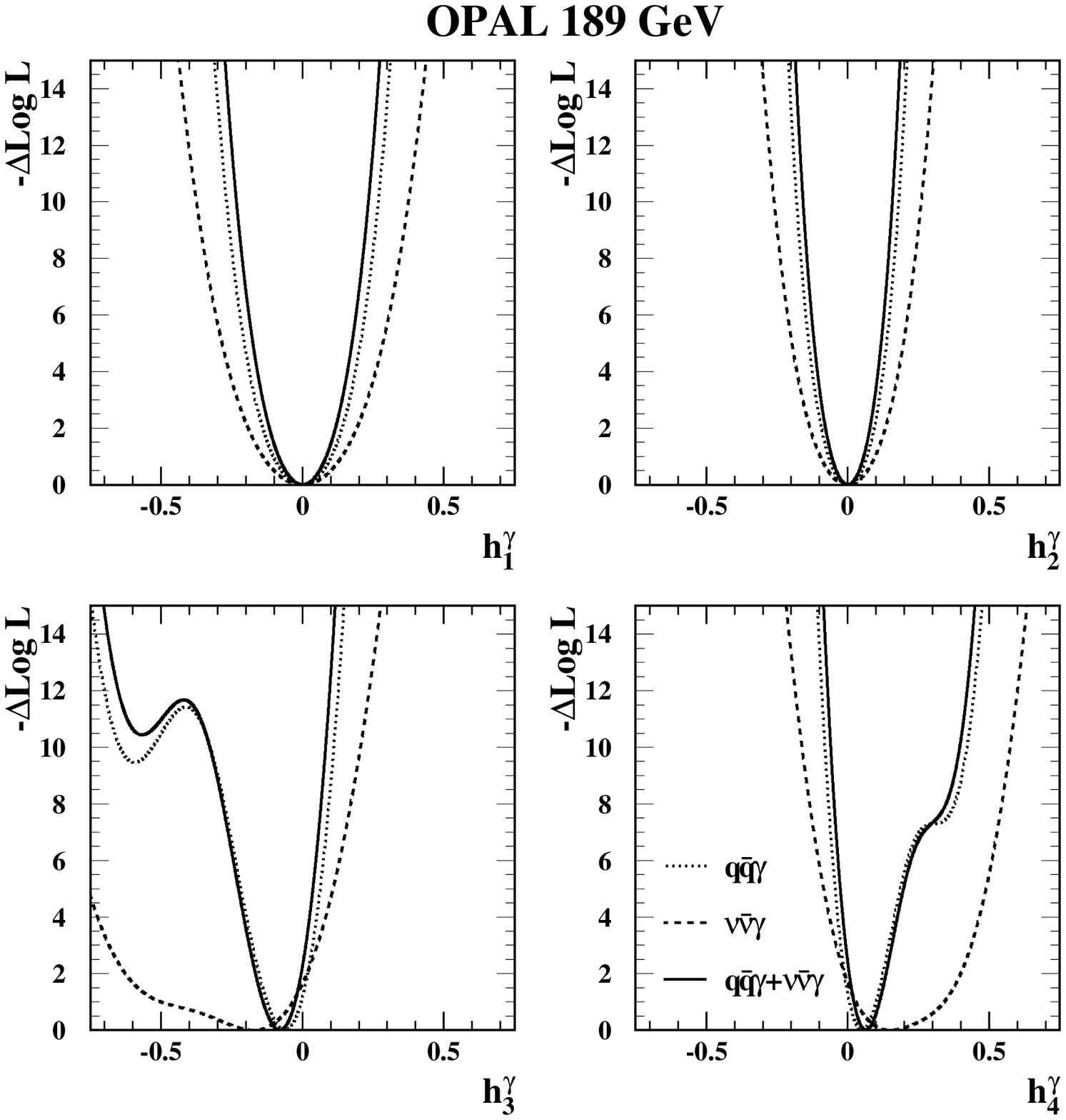}
  \caption
{Negative log-likelihood functions for the ${\rm h_{\it i}^{\gamma}}$ 
 couplings
  as obtained from the analysis of the \qqbarg channel 
 (dash-dotted line), of the
  $\nu\bar{\nu}\gamma$ channel (dashed line) 
  and from their combination (solid line).}
\label{fig:likg}
\end{figure}

The 95\% Confidence Level (C.L.) bounds on the eight 
anomalous couplings have been obtained by
convolving the likelihood function with a 
Gaussian whose width $\sigma$  corresponds to the systematic
uncertainty on the individual parameters, as estimated in 
section~\ref{sec:interpretation}.3.
The central values and the 95\% C.L. intervals 
resulting from the combination 
of the two channels are given in table \ref{tab:convqqg}.
Figures \ref{fig:likz} and \ref{fig:likg} show the corresponding 
negative log-likelihood curves for the individual channels and 
their combination. 
In combining the results, the correlations between the 
systematic uncertainties have been taken into account.

\subsection{Systematic Errors}

The impact  of several sources of 
systematic uncertainty on the NTGC determination 
has been assessed. Most of the effects considered
have already been discussed in the context of the systematic error on the 
cross-section measurements.  A few more are related to the modelling of NTGC
effects on the observables used in the likelihood fit and to the reference
Standard Model predictions. 
For all the sources of systematic uncertainty, a symmetric
error is assigned based on the maximum absolute shift 
between the central value, the lower and the upper 
edges of the 68\% C.L. interval as obtained in the standard analysis 
(table \ref{tab:agc}) and in a specific fit 
to the data performed to simulate the systematic effect. 
The dominant sources of systematic uncertainty come from:

\begin{itemize}

\item[-] The effects of the modelling of the selection efficiency.
  These have been evaluated using the same methods as discussed in 
  section \ref{sec:sysxs},
  and  have been assigned as 
  scale uncertainties on 
  the expected number of events in each channel. 
  Where relevant, the angular dependence of 
  the uncertainties has been taken into account.
  For the \qqbarg channel,  
  this includes also the uncertainties 
  coming from  the jet multiplicity, the 
  hadronisation modelling and 
  the jet parameter smearing.
 
\item[-] The theoretical uncertainty on 
the Standard Model prediction.
Considerations about missing higher order corrections
in the \kk\  Monte Carlo lead to an
estimate of the theoretical uncertainty  of the order of 
  1\% in the \qqbarg channel \cite{bib:KK}.
In the \nunug channel, a 2\% theoretical uncertainty 
has been  assigned; this uncertainty covers 
the observed differences between the kinematic 
cross-sections and the selection efficiencies estimated 
by \KORALZ\ and \NUNUGPV, and is consistent with a recent 
comparison between different theoretical calculations
presented in~\cite{bib:grcnn}. 
These theoretical uncertainties have been assigned as an overall 
normalisation error on the Standard Model prediction in each channel. 
\end{itemize}
Other minor contributions to the systematic error have 
been considered: 

\begin{itemize}

\item[-] The contribution of the background to the fit result. This has been
  separated into a component due to a scale factor (corresponding to the
  Monte Carlo statistical uncertainty on the background absolute rate) and 
  a component due to the modelling of the background shape. 
  The error due to the background normalisation is evaluated as the maximum
  effect observed when increasing or decreasing the 
  total background rate by one standard deviation. The error due 
  to the shape modelling has been 
  conservatively assessed by assuming a flat distribution of the 
  background events in the signal acceptance region. In the \nunug channel, 
  due to the very small contamination, only the uncertainty on the 
  background rate has been considered. 

\item[-] The uncertainty related to the Monte Carlo statistics. This
  effect has been estimated by applying the likelihood fit to the 
  data using, as Standard Model prediction in each bin, a number of 
  events generated
  according to a Poisson distribution with average equal to the
  prediction of the reference Monte Carlo sample and rescaled to the
  integrated luminosity of the data. The maximum among the r.m.s 
 of the central values and of the lower and upper 68\% C.L. limits 
  on each coupling, over a thousand fits, has been 
  assigned as systematic error.

\item[-] The uncertainty on the reweighting procedure 
coming from the limited Monte Carlo statistics 
used to evaluate the corrections for resolution and selection efficiency,
and from the limited Monte Carlo statistics  
used in the calculation of the generator-level weights.

\item[-]  
The uncertainty on the missing t-channel W exchange contribution, 
which applies only to the 
\nunug channel. The procedure 
used to correct for the missing W t-channel 
in  the reweighting procedure
accounts for this process only in the 
pure Standard Model contribution, but not in the 
interference between the Standard Model process and 
the process leading to anomalous coupling.
Since the W t-channel  is calculated to 
contribute less than 4\% to the 
Standard Model cross-section in the signal acceptance, the effect of it being 
neglected in the interference is expected to be of order 2\%.
The interference term has been conservatively varied by $\pm 4\%$
and the differences in the fit results have been assigned as a 
systematic error.

\item[-] The  systematic errors arising from the calorimeter
 energy scale and resolution, the modelling of the 
$\theta_\gamma$ angular cut, the uncertainty on the beam energy 
and on the integrated luminosity. They have been evaluated 
with the same method as discussed in section \ref{sec:sysxs}
 and have been treated as fully correlated between the 
\qqbarg channel and the \nunu$\gamma$ channel.
%, although their
%relative importance with respect to the  
%statistical error is, in general, different in the 
%two channels.

\end{itemize}

The systematic errors on the couplings are summarised 
in table \ref{tab:sysntgc} 
for both the \qqbarg  channel and the \nunug channel.
For all the couplings the total systematic error is small 
compared to the statistical uncertainty, except 
for $\rm h_{3,4}^\gamma$, where the size of 
the systematic error reaches 60\% of the statistical error. 

%\subsection{95\% C.L. Bounds on the Anomalous Couplings}

%%-----------------------------------------------------------------------
\section{Conclusions}    \label{sec:sum}

Using the data collected at $\sqrt{s}=189$~\GeV,
the cross-sections and the differential distributions for
hadronic events with a high energy isolated photon and 
for events with an energetic photon and missing energy
have been measured to search for 
possible contributions from 
anomalous ${\rm Z\gamma Z}$ and ${\rm Z\gamma\gamma}$ couplings.  
Since no significant evidence of deviations with respect to the 
Standard Model is observed, 
 95\% C.L. limits on the eight trilinear neutral gauge couplings 
$\rm{h_{\it i}^{Z,\gamma}}$ have been derived. 
These limits do not yet allow to place significant constraints
on specific models of new physics~\cite{bib:gounren}
leading to effective anomalous couplings in the neutral sector. Nevertheless, 
these results and those presented 
in~\cite{bib:ntgcL3}, which are compatible and of equivalent sensitivity,
represent the best available investigations 
in $\rm{e^+e^-\rightarrow Z\gamma}$ of the 
neutral gauge boson self-interactions, in the 
recently revised theoretical framework describing the general 
$\rm{Z\gamma V~~(V=Z,\gamma)}$ vertex.

%Using the data collected at $\sqrt{s}=189$~\GeV\,
%signals of anomalous ${\rm Z\gamma Z}$ and ${\rm Z\gamma\gamma}$
%couplings  have been searched for 
%in hadronic events with a high energy isolated photon and in
%events with an energetic photon and missing energy. 
%The cross-sections for \qqbarg and \nunug 
%production have been measured within a 
%specific signal definition, designed to enhance the sensitivity 
%to possible contributions from anomalous couplings.  
%%The measurements are in reasonable agreement with the Standard Model 
%%prediction.  
%Since no significant evidence of deviations with respect to the 
%Standard Model is observed, the measured cross-sections and differential 
%distributions have been used to derive 95\% C.L. 
%limits on the eight trilinear neutral gauge couplings 
%leading to anomalous Z$\gamma$ production. 
%The results are compatible with those presented in~\cite{bib:ntgcL3}.

%%-----------------------------------------------------------------------

%\input{acknowledgements.tex}
\section*{Acknowledgements}
The authors wish to thank U. Baur, G.J. Gounaris and F.M. Renard 
for helpful discussions and clarifications.
We particularly wish to thank the SL Division for the efficient operation
of the LEP accelerator at all energies
 and for their continuing close cooperation with
our experimental group.  We thank our colleagues from CEA, DAPNIA/SPP,
CE-Saclay for their efforts over the years on the time-of-flight and trigger
systems which we continue to use.  In addition to the support staff at our own
institutions we are pleased to acknowledge the  \\
Department of Energy, USA, \\
National Science Foundation, USA, \\
Particle Physics and Astronomy Research Council, UK, \\
Natural Sciences and Engineering Research Council, Canada, \\
Israel Science Foundation, administered by the Israel
Academy of Science and Humanities, \\
Minerva Gesellschaft, \\
Benoziyo Center for High Energy Physics,\\
Japanese Ministry of Education, Science and Culture (the
Monbusho) and a grant under the Monbusho International
Science Research Program,\\
Japanese Society for the Promotion of Science (JSPS),\\
German Israeli Bi-national Science Foundation (GIF), \\
%Bundesministerium f\"ur Bildung, Wissenschaft,
%Forschung und Technologie, Germany, \\
Bundesministerium f\"ur Bildung und Forschung, Germany, \\
National Research Council of Canada, \\
Research Corporation, USA,\\
Hungarian Foundation for Scientific Research, OTKA T-029328,
T023793 and OTKA F-023259.

\appendix

%%%%%%%%%%%%%%%%%%%%%%%%% References %%%%%%%%%%%%%%%%%%%%%%%%%%%%%

%%-----------------------------------------------------------------------
%%-----------------------------------------------------------------------
%%---------- here all the tables --------------------------------------%%
%%-----------------------------------------------------------------------
%%-----------------------------------------------------------------------
\clearpage
\begin{table}[htbp]
\centering
\begin{tabular}{|c|c|c|}
\hline%-----------------------------------------------------------------
\hline%-----------------------------------------------------------------
\multicolumn{3}{|c|}{\bf Efficiencies and Feedthrough}
\\
\hline%-----------------------------------------------------------------
\hline%-----------------------------------------------------------------
 Channel & Efficiency (\%)  & Feedthrough (\%)  \\ 
\hline%-----------------------------------------------------------------
 \qqbarg &  $87.58 \pm 0.12$ & $2.21 \pm 0.07$ \\
 \nunug  &  $81.04 \pm 0.22$ & $1.80 \pm 0.08$  \\
%\hline%-----------------------------------------------------------------
\hline%-----------------------------------------------------------------
\hline%-----------------------------------------------------------------
\end{tabular}
\caption[]{
  The  efficiency and the feedthrough 
  for the \qqbarg and \nunug selections.
  In the \nunug channel, they are the averages of the  
  KORALZ and the NUNUGPV Monte Carlo.
  The errors are from Monte Carlo statistics.}
\label{tab:eff}
\end{table}
%------------------------------------------------------------
\begin{table}[htbp]
\centering
\begin{tabular}{|c|c|c|c|c|c|}
\hline%-----------------------------------------------------------------
\hline%-----------------------------------------------------------------
\multicolumn{6}{|c|}{\bf Selected Events} 
\\
\hline%-----------------------------------------------------------------
\hline%-----------------------------------------------------------------
%{\footnotesize
  Observed &  \qqbar$\gamma$ SM & Four-Fermion
    & Two-Photon & $\tau^+\tau^-$ & Total
         expected \\ 
%}
\hline%-----------------------------------------------------------------
%{\footnotesize
 1525  &  $1538.7\pm 5.5$ &  $24.3 \pm 0.9$ & $9.0\pm 1.0$ &  $5.2 \pm  0.3$ & 
          $1577.2\pm 5.7$  \\
%}
\hline%-----------------------------------------------------------------
\hline%-----------------------------------------------------------------
\end{tabular}
\caption[]{The number of selected \qqbarg\ events in the data,
              compared with the 
              Monte Carlo
              expectations for the signal and for the 
              background, normalised to the data
              integrated luminosity. The errors are from 
              Monte Carlo statistics.
              \label{tab:events}}
\end{table}
%-----------------------------------------------------------------
\begin{table}[htbp]
\centering
\begin{tabular}{|c|c|c|c|c|}
\hline%-----------------------------------------------------------------
\hline%-----------------------------------------------------------------
\multicolumn{5}{|c|}{\bf Selected Events} 
\\
\hline%-----------------------------------------------------------------
\hline%-----------------------------------------------------------------
%{\footnotesize
 Observed & \nunug SM & Physics Background &  Instrumental Background & 
 Total expected\\ 
%}
\hline%-----------------------------------------------------------------
              370 &  
              $411.6 \pm 2.5$ & 
              $0.72 \pm  0.83$ & 
              $0.18\pm 0.18$  &
              $412.5\pm 2.6$  \\
\hline%-----------------------------------------------------------------
\hline%-----------------------------------------------------------------
\end{tabular}
\caption[]{The number of \nunug events selected in 
data,  the expectation from the \nunug Standard Model process, 
from physics processes 
other than \nunug, the expected contamination from cosmic rays 
and beam-related backgrounds and the overall number of expected events
in the \nunug channel. The uncertainties 
reflect either  the Monte Carlo statistics or the statistics 
of the control samples used for the instrumental background estimate.}
\label{tab:eventsnng}
\end{table}
\begin{table}[h]
\centering
\begin{tabular}{|c|c|c|c|}
\hline%-----------------------------------------------------------------
\hline%-----------------------------------------------------------------
\multicolumn{4}{|c|}{\bf Systematic errors on the cross section}\\
\hline%-----------------------------------------------------------------
{\bf \qqbarg } channel & 
$\rm{\Delta\sigma}$ (pb) & 
{\bf \nunug} channel & 
$\rm{\Delta\sigma}$ (pb)\\
\hline%-----------------------------------------------------------------
\hline%-----------------------------------------------------------------
Efficiency                        &    0.115   & Efficiency& 0.045\\ 
Jet Modelling and Reconstruction  &    0.074   & - & - \\
Background                        &    0.066   & Background& 0.006 \\
\hline %----------------------------------------------------------------
$\theta_\gamma$ Cut               &    0.018   & $\theta_\gamma$ Cut & 0.005\\
Energy Scale                      &    0.009  & Energy Scale & 0.003\\  
Energy Resolution                 &    0.008  & Energy Resolution & 0.012\\
Beam Energy                       &    0.003  & Beam Energy & 0.001\\
Luminosity                        &    0.020  & Luminosity & 0.005\\ 
\hline %----------------------------------------------------------------
\hline %----------------------------------------------------------------
   Total Systematic Error            &    0.154  
&  Total Systematic Error            &    0.048 \\ 
\hline %----------------------------------------------------------------
\hline%-----------------------------------------------------------------
\end{tabular}
\caption[]{
Systematic errors on the measurement of
the cross-section for the \qqbarg and \nunug channels.}
\label{tab:sysxs}
\end{table}%------------------------------------------------------------
\begin{table}[htbp]
\centering
\begin{tabular}{|c|cc|cc|}
\hline%-----------------------------------------------------------------
\hline%-----------------------------------------------------------------
\multicolumn{5}{|c|}{\bf NTGC Fit Results and Expected Sensitivity}\\
\hline%-----------------------------------------------------------------
    & ~~~~~~~~~~~~~\qqbarg  & channel~~~~~~~~~~~~~ & ~~~~~~~~~~~~~\nunug  & channel~~~~~~~~~~~~~ \\ 
\hline%-----------------------------------------------------------------
\hline%-----------------------------------------------------------------
Coupling & Fit Results  & Sensitivity & Fit Results  & Sensitivity \\ 
\hline%-----------------------------------------------------------------
 & & & & \\
 ${\rm h_1^Z}$     &
 $0.000\pm 0.122$  &  
 $\pm 0.26$        & 
 $0.000\pm 0.166$  & 
 $\pm 0.35$     \\
%---------------------------------------------------------------
 & & & & \\
 ${\rm h_2^Z}$     &
 $0.000\pm 0.081$  &  
 $\pm 0.17$        & 
 $0.000 \pm 0.115$ & 
 $\pm 0.25$     \\ 
%---------------------------------------------------------------
 & & & & \\
${\rm h_3^Z}$      &
 $-0.055^{+0.122}_{-0.126}$ &
 $-0.33, +0.19$    & 
 $-0.107 \pm 0.169$ & 
 $-0.48,\: +0.37$ \\ 
%---------------------------------------------------------------
 & & & & \\
${\rm h_4^Z}$      & 
$0.035^{+0.082}_{-0.081}$ & 
$-0.14,\: +0.21$  & 
$0.067 \pm 0.116$ & 
$-0.19,\: +0.25$ \\ 
 & & & & \\
\hline%-----------------------------------------------------------------
 & & & & \\
${\rm h_1^\gamma}$ 
& $0.000\pm 0.074$ 
& $\pm 0.15$ 
& $0.000\pm 0.100$ 
& $\pm 0.21$ \\
%-----------------------------------------------------------------
 & & & & \\
${\rm h_2^\gamma}$ 
& $0.000\pm 0.049$ 
&  $\pm 0.10$
& $0.000\pm 0.069$ 
& $\pm 0.15$ \\ 
%-----------------------------------------------------------------
 & & & & \\
${\rm h_3^\gamma}$ 
& $-0.061^{+0.035}_{-0.038}$ 
& $-0.033,\:+0.031$ 
& $-0.163^{+0.087}_{-0.139}$ 
& $-0.068,\: +0.064$ \\ 
%-----------------------------------------------------------------
 & & & & \\
${\rm h_4^\gamma}$ 
& $ 0.049^{+0.031}_{-0.027}$ 
& $-0.023,\: +0.025$ 
& $0.138^{+0.143}_{-0.076}$ 
& $-0.045,\: +0.050$  \\
 & & & & \\
\hline%-----------------------------------------------------------------
\hline%-----------------------------------------------------------------
\end{tabular}
\caption[]{
Best fit values of the anomalous couplings and the corresponding 
 statistical errors for the \qqbarg and \nunug channels. 
 The expected statistical errors, obtained 
 under the hypothesis of 
 perfect agreement with the Standard Model expectations, 
 are also shown.}
\label{tab:agc}
\end{table}%------------------------------------------------------------

\begin{table}[htbp]
\centering
\begin{tabular}{|c|c|c|}
\hline%-----------------------------------------------------------------
\hline%-----------------------------------------------------------------
\multicolumn{3}{|c|}{\bf Combined Results on NTGC}
\\
\hline%-----------------------------------------------------------------
\hline%-----------------------------------------------------------------
Coupling & Central Value & 95\% C.L.  Interval~~~ \\
\hline%-----------------------------------------------------------------
\hline%-----------------------------------------------------------------
 & & \\
${\rm h_1^Z}$ & 0.000$\pm$0.100            & [~$-0.190,~+0.190$~] \\
 & & \\
${\rm h_2^Z}$ & 0.000$\pm$0.068            & [~$-0.128,~+0.128$~] \\
 & & \\
${\rm h_3^Z}$ & $-0.074^{+0.102}_{-0.103}$ & [~$-0.269,~+0.119$~] \\
 & & \\
${\rm h_4^Z}$ & $ 0.046^{+0.068}_{-0.068}$ & [~$-0.084,~+0.175$~] \\
 & & \\
\hline%-----------------------------------------------------------------
 & & \\
${\rm h_1^\gamma}$ & 0.000$\pm$0.061 & [~$-0.115,~+0.115$~] \\
 & & \\
${\rm h_2^\gamma}$ & 0.000$\pm$0.041 & [~$-0.077,~+0.077$~] \\
 & & \\
${\rm h_3^\gamma}$ & $-0.080^{+0.039}_{-0.041}$ & [~$-0.164,~-0.006$~] \\
 & & \\
${\rm h_4^\gamma}$ & $ 0.064^{+0.033}_{-0.030}$ & [~$+0.007,~+0.134$~] \\
 & & \\
\hline%-----------------------------------------------------------------
\hline%-----------------------------------------------------------------
\end{tabular}
\caption{ Central values and 95\% Confidence Level intervals for the eight
  anomalous couplings as determined from the combination of the 
  \qqbarg and the \nunug channels. Systematic uncertainties have been 
  incorporated in the errors and in the 95\% C.L. limits.}
\label{tab:convqqg}
\end{table}
\begin{table}[htbp]
\centering
\begin{tabular}{|c|c|c|}
\hline%-----------------------------------------------------------------
\hline%-----------------------------------------------------------------
\multicolumn{3}{|c|}{\bf Systematic errors on NTGC}\\
\hline%-----------------------------------------------------------------
               &
{\bf \qqbarg} channel & 
{\bf \nunug} channel \\
\hline%-----------------------------------------------------------------
\hline%-----------------------------------------------------------------
${\rm \Delta h^{Z}_{1}}$ & 0.015 & 0.021 \\
${\rm \Delta h^{Z}_{2}}$ & 0.009 & 0.015\\
${\rm \Delta h^{Z}_{3}}$ & 0.023 & 0.025\\
${\rm \Delta h^{Z}_{4}}$ & 0.013 & 0.021\\
${\rm \Delta h^{\gamma}_{1}}$ & 0.009 & 0.012 \\
${\rm \Delta h^{\gamma}_{2}}$ & 0.006 & 0.009\\
${\rm \Delta h^{\gamma}_{3}}$ & 0.021 & 0.050\\
${\rm \Delta h^{\gamma}_{4}}$ & 0.016 & 0.044\\
\hline %----------------------------------------------------------------
\hline%-----------------------------------------------------------------
\end{tabular}
\caption[]{ Absolute systematic errors on
 the NTGC couplings determined
 in the \qqbarg\ and in the \nunug channel.}
\label{tab:sysntgc}
\end{table}%------------------------------------------------------------

%-----------------------------------------------------------------------
%       Figures
%-----------------------------------------------------------------------
%
\clearpage

%\input{agc_fig.tex}
%%%%%%%%%%%%%%%%%%%%%%%%%%%%%%%%%%%%%%%%%%%%%%%%%%%%%%%%%%
%%%%%%%%%%%%%%%%%%%%%%%%%%%%%%%%%%%%%%%%%%%%%%%%%%%%%%%%%%
%%%%%%%%%%%%%%%%%%%%%%%%%%%%%%%%%%%%%%%%%%%%%%%%%%%%%%%%%%
%%%%%%%%%%%%%%%%%%%%%%%%%%%%%%%%%%%%%%%%%%%%%%%%%%%%%%%%%%
%%%%%%%%%%%%%%%%%%%%%%%%%%%%%%%%%%%%%%%%%%%%%%%%%%%%%%%%%%
%%%%%%%%%%%%%%%%%%%%%%%%%%%%%%%%%%%%%%%%%%%%%%%%%%%%%%%%%%
%%%%%%%%%%%%%%%%%%%%%%%%%%%%%%%%%%%%%%%%%%%%%%%%%%%%%%%%%%
%%%%%%%%%%%%%%%%%%%%%%%%%%%%%%%%%%%%%%%%%%%%%%%%%%%%%%%%%%

\end{document}